\documentclass[fleqn,usenatbib]{mnras}

\usepackage{newtxtext,newtxmath}

\usepackage[T1]{fontenc}

\DeclareRobustCommand{\VAN}[3]{#2}
\let\VANthebibliography\thebibliography
\def\thebibliography{\DeclareRobustCommand{\VAN}[3]{##3}\VANthebibliography}

\usepackage{graphicx}	
\usepackage{amsmath}	

\title[Dark deflector bends nuclear jet]{IceCube AGN Neutrino candidate PKS\,1717+177: \\Dark deflector bends nuclear jet}

\author[S. Britzen et al.]{
S. Britzen,$^{1}$\thanks{E-mail: sbritzen@mpifr.de}
A.B. Kova\v{c}evi\'{c},$^{2}$
    M. Zaja\v{c}ek,$^{3}$
          L. \v{C}. Popovi\'{c},$^{2,4}$
          I.N. Pashchenko,$^{5,6,7}$
          E. Kun,$^{8,9,10,11,12}$
          \newauthor
          R. P\'{a}nis,$^{13}$
          F. Jaron,$^{14,1}$
    T. Pl\v{s}ek,$^{3}$
            A. Tursunov,$^{1,13}$
           and Z. Stuchl\'{i}k$^{13}$
\\
$^{1}$Max-Planck-Institut f\"ur Radioastronomie, Auf dem H\"ugel 69, 53 121 Bonn, Germany\\
$^{2}$Department of Astronomy, Faculty of Mathematics, University of Belgrade, Studentski trg 16,11000 Belgrade, Serbia\\
$^{3}$Department of Theoretical Physics and Astrophysics, Faculty of Science, Masaryk University, Kotl\'a\v{r}sk\'a 2, 611 37 Brno, Czech Republic\\
$^{4}$Astronomical Observatory Belgrade, Volgina 7, P.O.Box 74 11060, Belgrade, 11060, Serbia \\
$^{5}$Astro Space Center, Lebedev Physical Institute, Russian Academy of Sciences \\
$^{6}$Moscow Institute of Physics and Technology, Dolgoprudny, Institutsky per., 9, 141700 Moscow, Russia \\
$^{7}$Institute for Nuclear Research of the Russian Academy of Sciences, 60th October Anniversary Prospect 7a, Moscow 117312, Russia\\
$^{8}$Theoretical Physics IV: Plasma-Astroparticle Physics, Faculty for Physics \& Astronomy, Ruhr University Bochum, 44780 Bochum, Germany\\
$^{9}$Ruhr Astroparticle And Plasma Physics Center (RAPP Center), Ruhr-Universität Bochum 44780 Bochum, Germany\\
$^{10}$Astronomical Institute, Faculty for Physics \& Astronomy, Ruhr University Bochum, 44780 Bochum, Germany\\
$^{11}$Konkoly Observatory, HUN-REN Research Centre for Astronomy and Earth Sciences,
H-1121 Budapest, Konkoly Thege Miklós út 15-17., Hungary\\
$^{12}$CSFK, MTA Centre of Excellence, Konkoly Thege Miklós út 15-17., Hungary\\
$^{13}$Research Centre for Theoretical Physics and Astrophysics, Institute of Physics, Silesian University in Opava, Bezru\v{c}ovo n\'{a}m. 13, CZ-74601 Opava, Czech Republic\\       
$^{14}$Technische Universit\"at Wien, Wiedner Hauptstra\ss{}e 8-10, 1040 Wien, Austria\\
}

\date{Accepted XXX. Received YYY; in original form ZZZ}

\pubyear{2015}

\begin{document}
\label{firstpage}
\pagerange{\pageref{firstpage}--\pageref{lastpage}}
\maketitle

\begin{abstract}
The BL Lac Object PKS 1717+177 has been identified as potential neutrino-emitting AGN in the point source stacking analysis
of IceCube data. We explore peculiarities in the morphology and kinematics of the jet and examine multi-wavelength light
curves for distinctive effects which might allow to pinpoint a likely neutrino generation mechanism. We re-modeled 34 high
resolution radio interferometric Very Long Baseline Array (VLBA) observations obtained at 15 GHz (between 1999/12/27 and
2023/05/03). A correlation and periodicity analysis of optical KAIT and Tuorla data, as well as for \textit{Fermi}-LAT $\gamma$-ray data has
been performed. The nuclear jet appears deflected and bent at about 0.5 mas distance from the radio core by an encounter with a
dark, unseen object. The deviation of the jet evolves over 23.5 years from a simple apparent bend into a significantly meandering
structure with increasing amplitude: a zig-zag line. To our knowledge, this is the first time that the temporal evolution of a jet
deviation can be traced. The turning point shifts with time and the jet seems to brighten up almost periodically at the point of
deviation. The radio core as well as the jet contribute approximately equally to the total flux-density at 15 GHz. We discuss
scenarios which could explain the complex jet bending and quasi-regular flaring. We propose that the jet could either be deflected by the magnetosphere of a second massive black hole, by the pressure gradient due to a circumnuclear dense cloud, or via gravitational lensing by an intervening black hole.
\end{abstract}

\begin{keywords}
astroparticle physics -- black hole physics -- neutrinos -- gravitational lensing: strong -- BL Lacertae objects: individual: PKS\,1717+177 -- galaxies: active
\end{keywords}


\section{Introduction}
IceCube detects neutrinos of cosmic origin. However, so far it is not clear, from which source class these neutrinos predominantly originate. Moreover, the neutrino generating mechanism is not known either. A correlation is expected with $\gamma$-ray emitting blazars (e.g., \citealt{plavin}). The \textit{Fermi}-LAT routinely monitors 2863 blazars \citep{ajello}. Yet only a minority of them appear as likely candidates for neutrino emission (e.g., \citealt{garrappa}). 

While the jet is supposed to be a likely site of the neutrino emission, not much is known about the details of these emission processes. 
A recent stacking analysis of ten years of IceCube data has provided a list of potential AGN that might have generated neutrinos \citep{aartsen}. From this list, we selected blazars for a detailed investigation. In this manuscript, we study the low-synchrotron peak (LSP; $<$10$^{14}$~Hz) BL Lac object PKS\,1717+177. Further details concerning this source and the cosmological parameters applied in this manuscript are listed in Table \ref{list}.

The source was originally suggested as a BL Lac object because of its featureless optical spectrum by \citet{1976ApJS...31..143W}, confirmed later in \cite{1993A&AS..100..521V}. On kpc-scales it has a weak secondary component located 9.5 arcsec south of the core in radio observations at 1490 MHz \citep{1985ApJ...294..158A}. The source has a flat radio spectrum \citep[see references in][]{1993A&AS..100..521V}. \citet{2011ApJ...740...98M} classified the source as QSO and estimated the extended jet radio power $L_{\rm 300 MHz} = 1.26 \times 10^{40}$ erg/s using both the spectral decomposition and the extraction of the beamed core emission from the jet extended structure, that puts it on the borderline of FRI-FRII division \citep{1974MNRAS.167P..31F,1994ASPC...54..319O}.


 We perform a detailed analysis of high resolution radio interferometry observations available within the Monitoring Of Jets in Active galactic nuclei with VLBA Experiments (MOJAVE\footnote{\url{https://www.cv.nrao.edu/MOJAVE/}}) webpage. We combine this morphological investigation of the pc-scale jet structure with a periodicity, a correlation, and a nonlinear analysis of optical KAIT and Tuorla data, as well as of \textit{Fermi}-LAT $\gamma$-ray data. 
 
 \subsection{Neutrino-source candidates emerged from the time-integrated analysis of 10-years of IceCube data}
The time-integrated analysis with $10$ years of IceCube data covers the time between 2008/04/06 and 2018/07/10 \citep{aartsen}. The analysis targets astrophysical muon neutrinos and anti-neutrinos which undergo charged current interactions in the Antarctic ice to produce muons traversing the IceCube detector. The background measured by the detector is mostly from atmospheric cosmic rays that produce showers of particles, including muons and neutrinos. 

Source-catalog searches are conducted to improve sensitivity to detect possible neutrino sources in the 10-years neutrino dataset that might be observed at $\gamma$-rays. The extragalactic sources are taken from the $8$ years \textit{Fermi}-LAT 4FGL catalog \citep{Abdollahi2020}, while the Galactic sources are selected from TeVCat \citep{Wakely2008} and gammaCat catalogs. The spectra of $\gamma$-ray sources were converted to equivalent neutrino fluxes assuming a \textit{purely} hadronic origin of the observed $\gamma$-ray emission and compared to the sensitivity analysis at the declination of the source.

In the Northern hemisphere, $97$ neutrino source candidates ($87$ extragalactic and $10$ Galactic), while in the Southern hemisphere $14$ neutrino source candidates were found ($11$ extragalactic, $3$ Galactic). The large north-south difference is due to the difference in sensitivity of IceCube in the Northern and Southern hemispheres. The most significant excess in the Northern catalog is found in the direction towards NGC~1068, with $4.1\sigma$ pre-trial ($\hat{n}_s=50.4$ and $\hat{\gamma}=3.2$), and $2.9\sigma$ post-trial probabilities against the background only null-hypothesis. The hottest spot in the Southern hemisphere was found in the direction of PKS\,2233-148, with $0.06$ pre-trial and $0.55$ post-trial probability, which is consistent with the background.

From the population study, a post-trial $p$-value of $4.8\times 10^{-4}$ emerged for the Northern catalog of $97$ objects, that provides a $3.3\sigma$ inconsistency with a background-only hypothesis for the catalog. This is mainly due to the excess of significant $p$-values in the directions of NGC\,1068, TXS\,0506+056, PKS\,1424+240 and GB6\,J1542+6129 (all four found with pre-trial $p$-value $<0.01$). 

The main source of this paper, PKS\,1717+177, is a BL Lac object in the Northern catalog, and a neutrino source candidate. The best-fit number of astrophysical neutrino events emerged for this source as $\hat{n}_s=19.8$, the best-fit astrophysical power-law spectral index as $\hat{\gamma}=3.6$, and the local pre-trial as $p$-value $=0.05$.

The paper is organized as follows. In Section~\ref{sec_obs}, we first present the applied observations and data reduction techniques. In Section~\ref{sec_results}, we then focus on the results concerning the pc-scale jet dynamics, as well as a correlation and periodicity analysis of single-dish radio, optical data (KAIT and Tuorla), and the \textit{Fermi}-LAT light curve. We discuss our results in view of the peculiar properties found and their relevance for the neutrino production mechanisms in Section~\ref{sec_discussion} and briefly summarize our findings in Section~\ref{sec_conclusions}.



 \begin{figure*}
   \centering
 \begin{minipage}{0.45\textwidth}
    \centering
    \includegraphics[width=\linewidth]{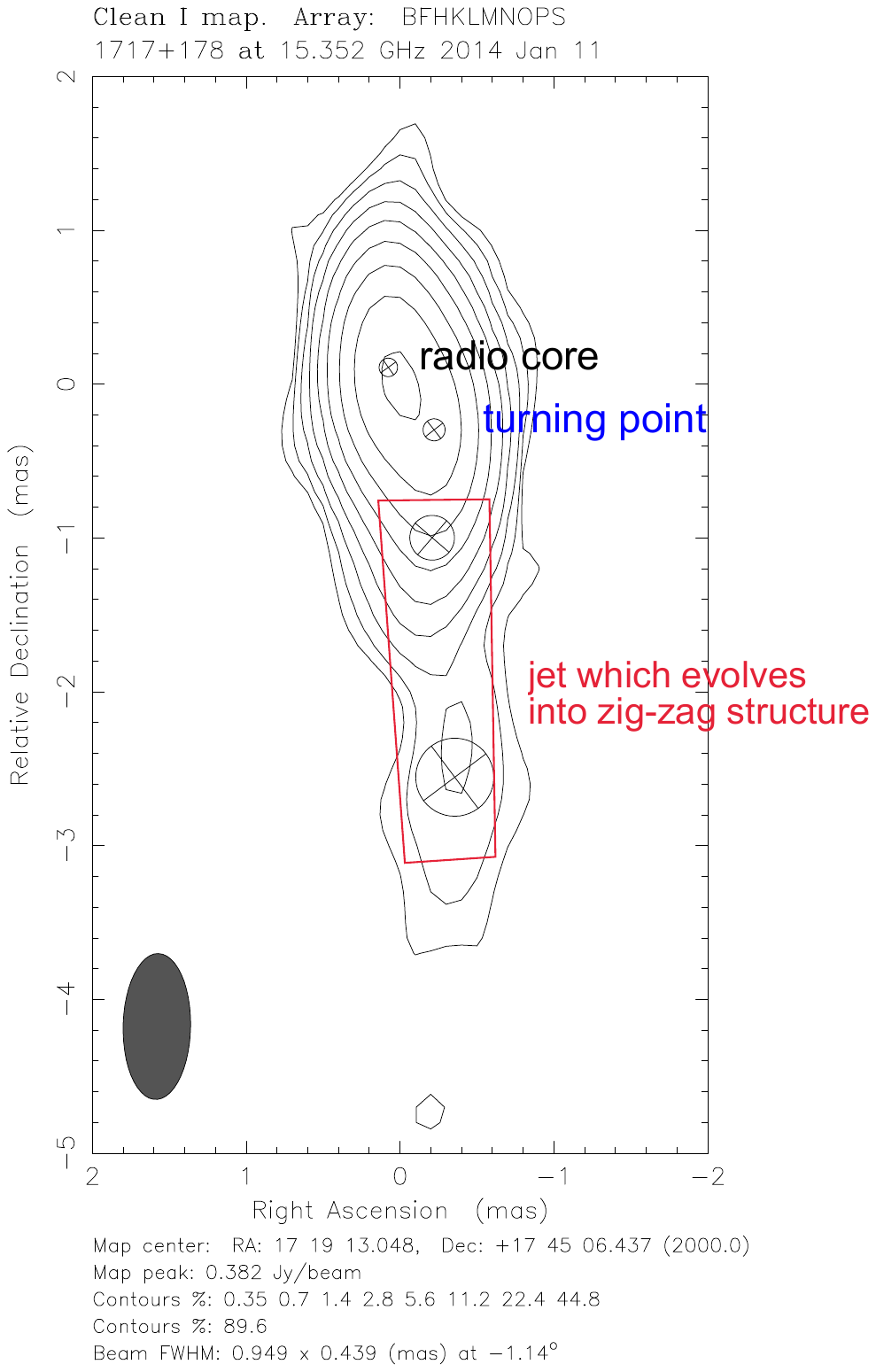}
    [a]
\end{minipage}
\begin{minipage}{0.45\textwidth}
    \centering
    \includegraphics[width=\linewidth]{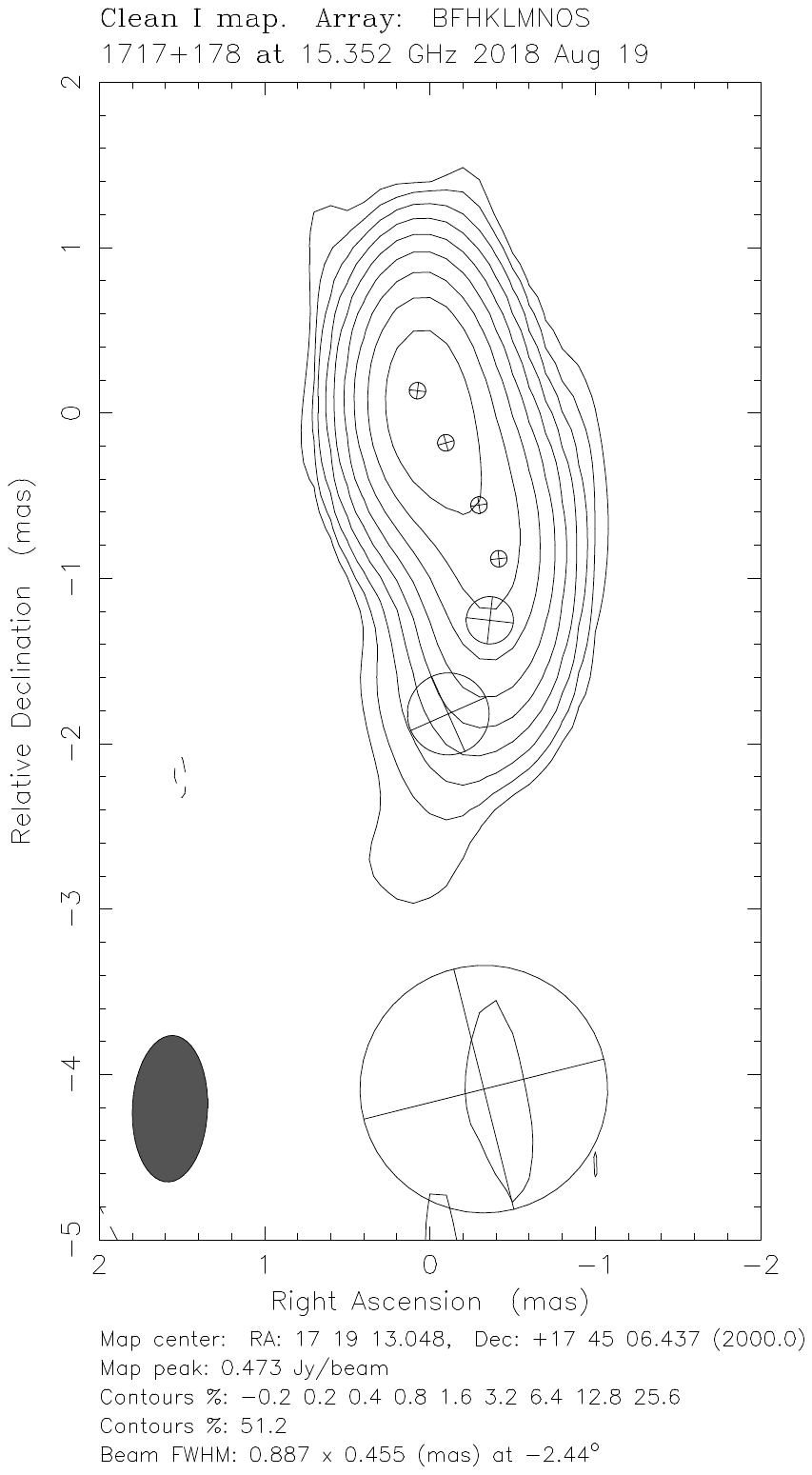}
    [b]
\end{minipage}
   \caption{Gaussian modelfit components are superimposed on a map of PKS\,1717+177 at epoch 11/01/2014 (a) and epoch 19/08/2018 (b). The main features discussed further in the text, are annotated in (a). Obviously, the jet looks different at both epochs. We trace the continuous evolution of the jet bending and discuss an encounter with a dark deflector as possible cause in the text.}%
   \label{maps}
    \end{figure*}

     \begin{table*}
      \caption[]{We collected information on the type, redshift, detection in the $\gamma$-ray and TeV-regime from the MOJAVE webpage. The following cosmological parameters are used: $H_{0} = 71 {\rm km/s/Mpc}$, $\Omega_{\Lambda} = 0.73$ and $\Omega_{\rm m} = 0.27$. 
      }
         \label{list}
         \centering
         \begin{tabular}{llllcc}
            \hline
            \noalign{\smallskip}
      Source   & Type  & z & $\gamma$-ray emission& scale&Black Hole mass\\
            &&& &[pc/mas]&M$_\odot$\\
            \noalign{\smallskip}
            \hline
            \noalign{\smallskip}
       PKS\,1717+177, J1719+1745, OT 129&LSP BL Lac&0.137&LAT: Y, TeV: N&2.40&3$\times$10$^8$ \citep{ghisellini}\\
            \noalign{\smallskip}
            \hline
         \end{tabular}
         \label{tab_source}
   \end{table*} 
\begin{figure*}
   \centering
 \begin{minipage}{0.45\textwidth}
    \centering
    \includegraphics[width=\linewidth]{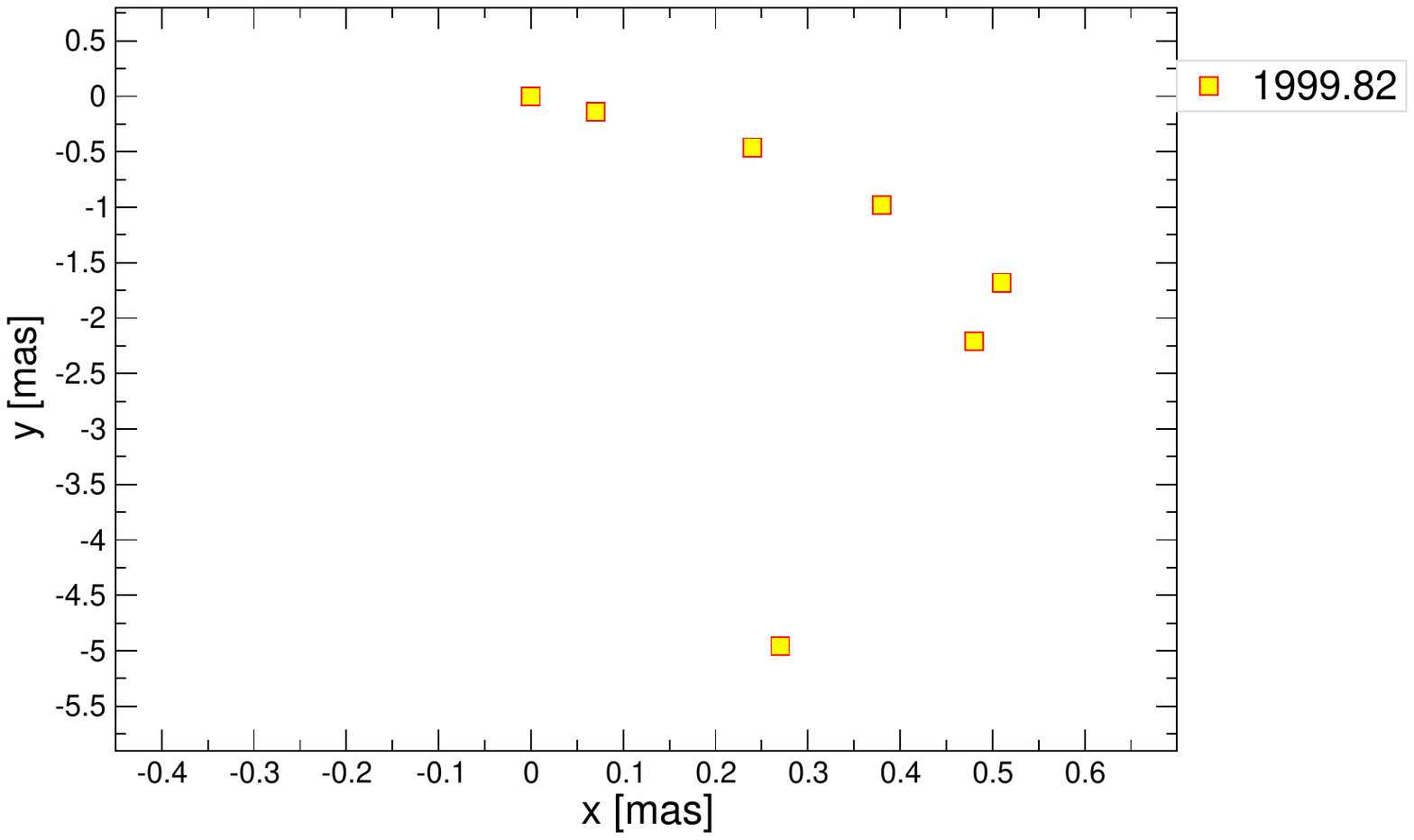}
    [a]
\end{minipage}
\begin{minipage}{0.45\textwidth}
    \centering
    \includegraphics[width=\linewidth]{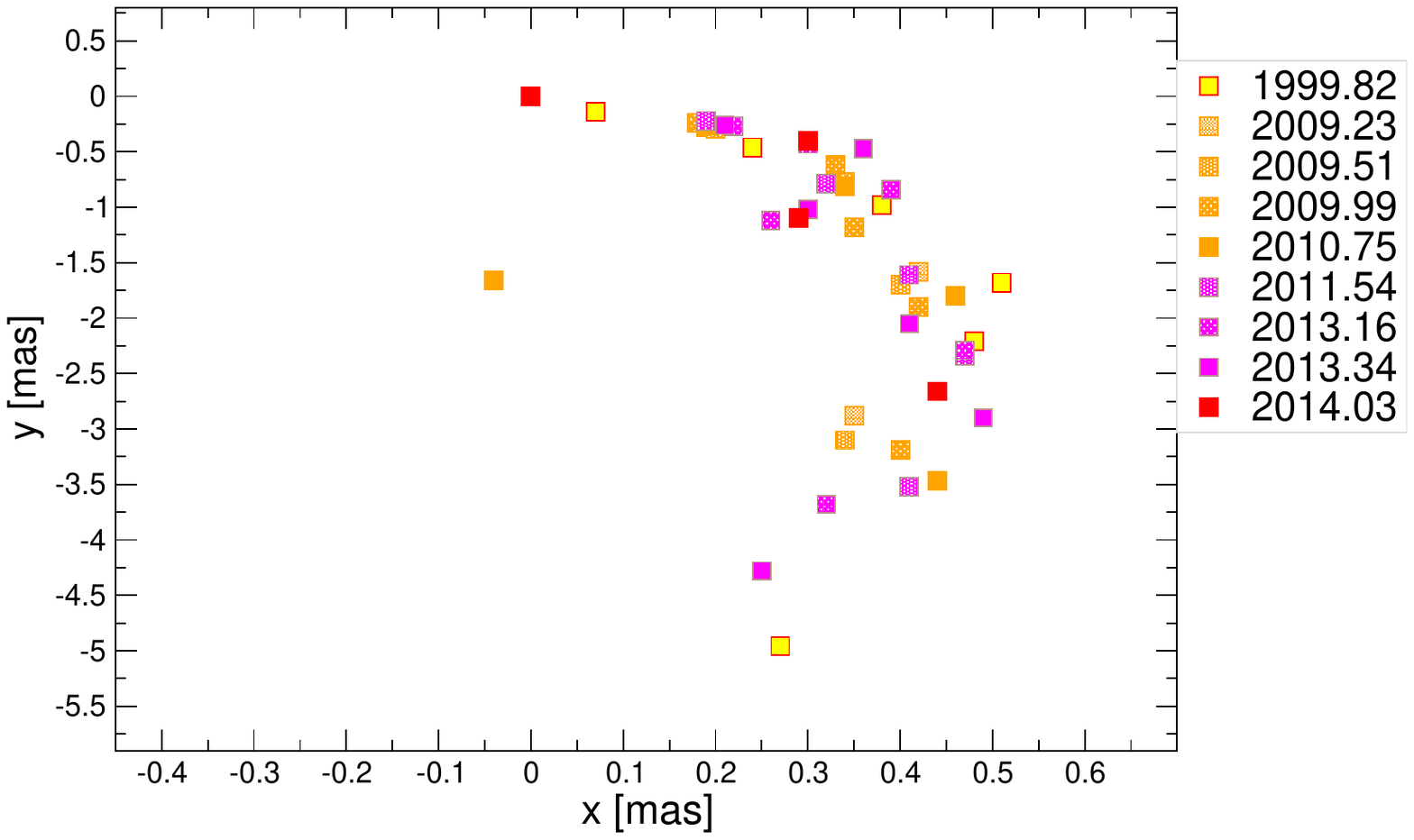}
    [b]
\end{minipage}
 \begin{minipage}{0.45\textwidth}
    \centering
    \includegraphics[width=\linewidth]{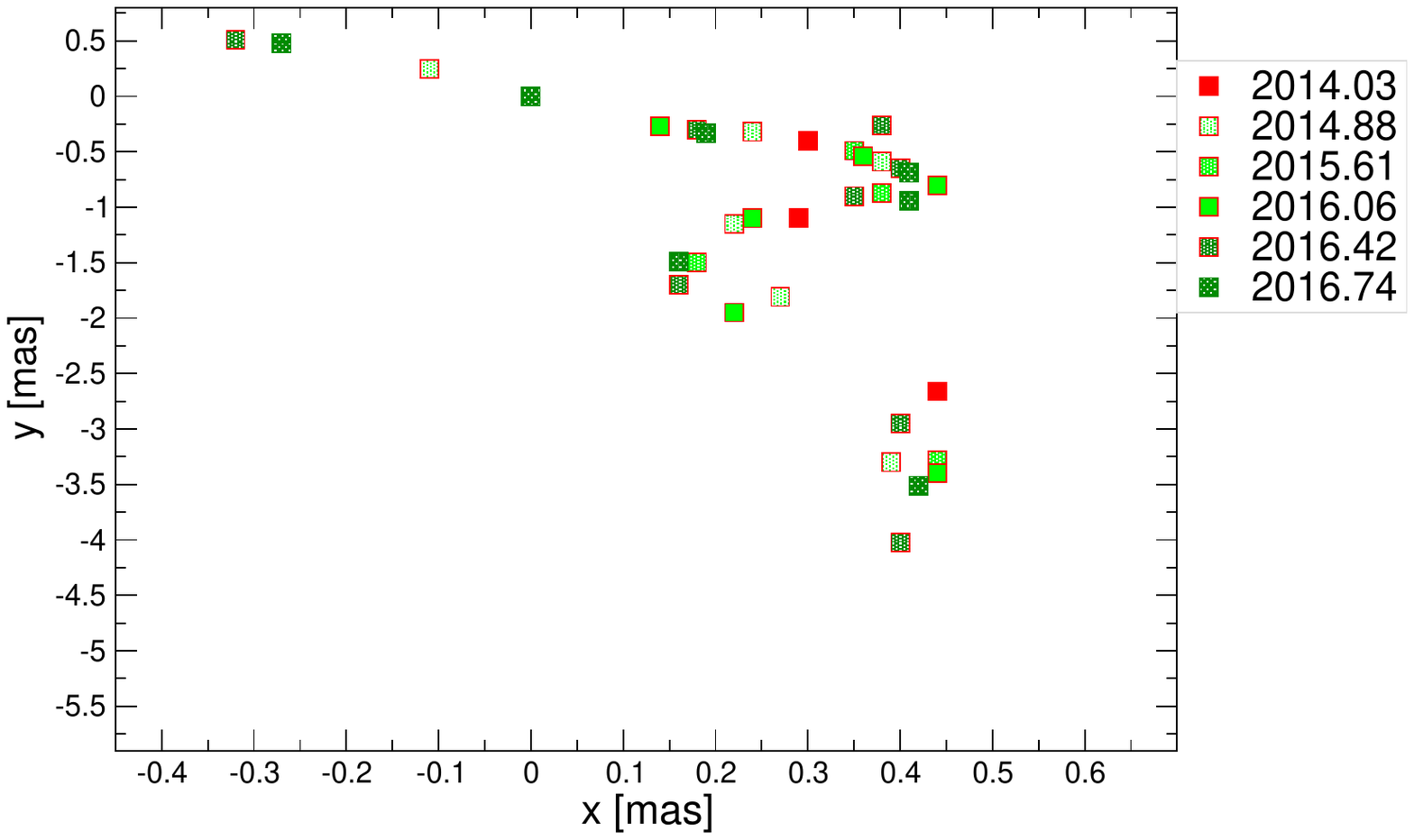}
    [c]
\end{minipage}
\begin{minipage}{0.45\textwidth}
    \centering
     \includegraphics[width=\linewidth]{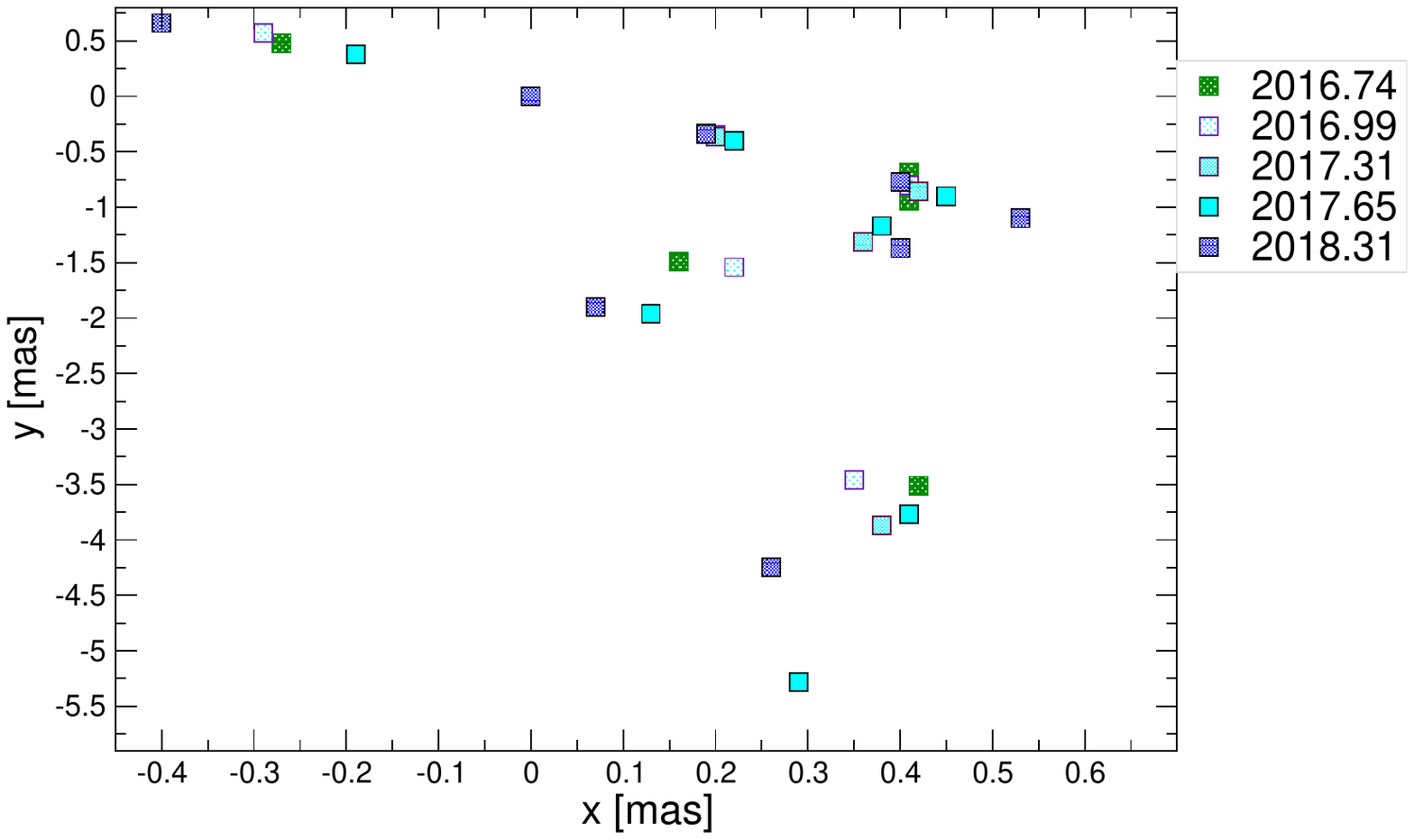}
    [d]
\end{minipage}
 \begin{minipage}{0.45\textwidth}
    \centering
    \includegraphics[width=\linewidth]{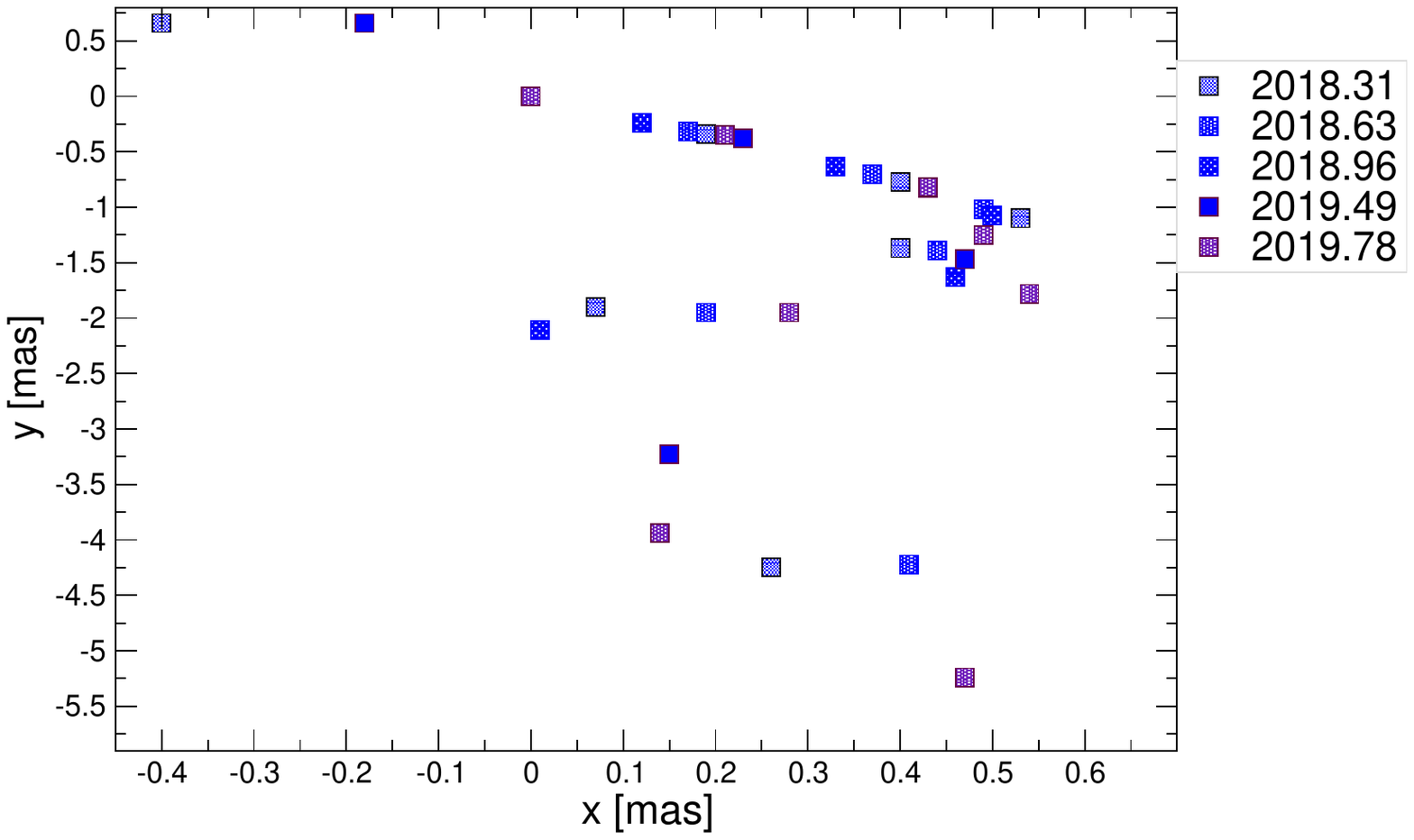}
    [e]
\end{minipage}
\begin{minipage}{0.45\textwidth}
    \centering
    \includegraphics[width=\linewidth]{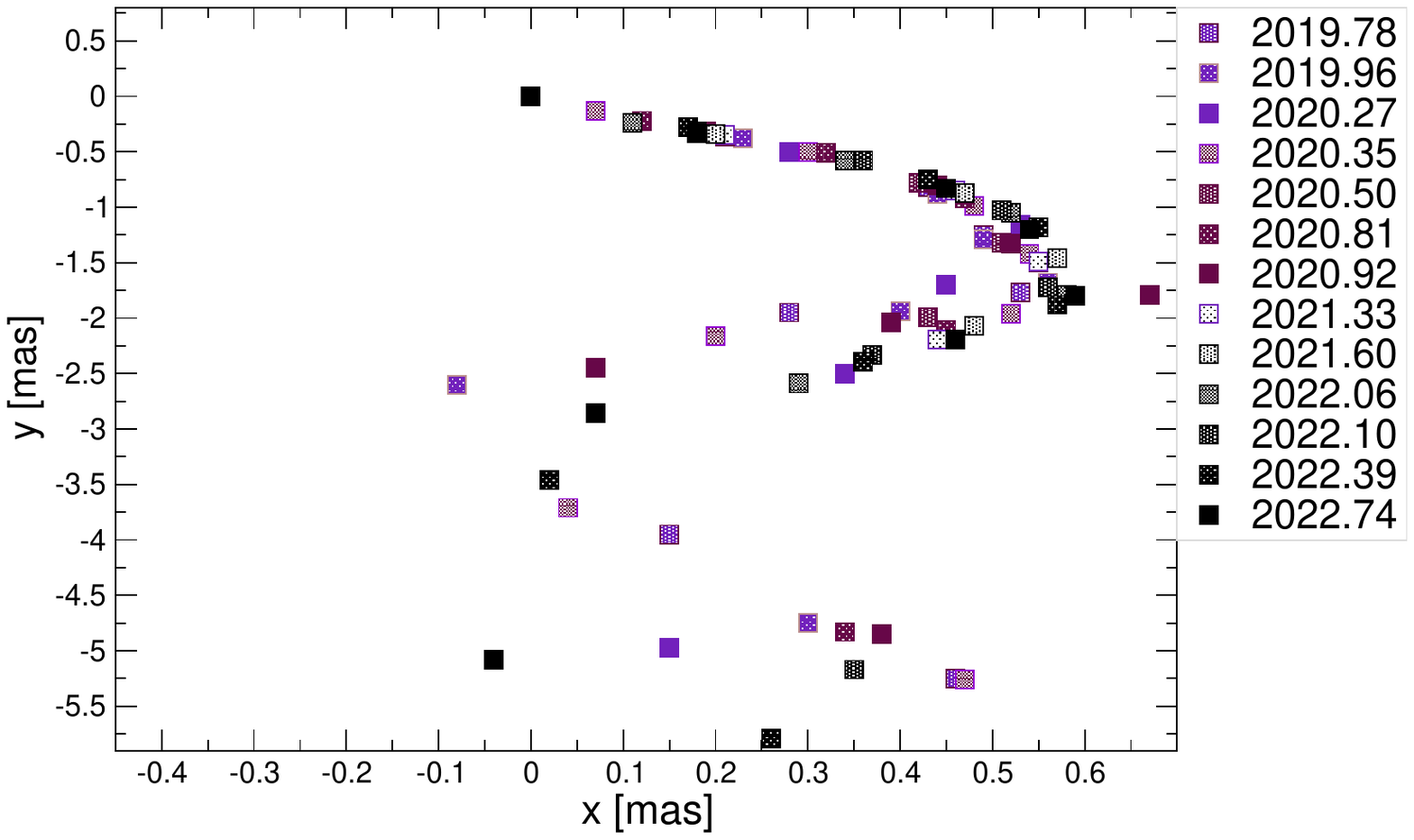}
    [f]
\end{minipage}
   \caption{The temporal evolution of the jet of PKS\,1717+177 is shown in [a] to [f] to highlight special phases of the bending. Two data points at large core separation (larger than 0.9 mas in x-coordinate) are not shown in these plots but are included in Fig.\,\ref{xy}. The curvature of the jet changes significantly and this evolution in time can be traced with the VLBA observations in xy-positions Please note the different scales on the x- and y-axis. For better visualization we chose to plot non-equal scales.}
   \label{xy_plots}
    \end{figure*}
   
\begin{figure}
   \centering
  \begin{minipage}{\columnwidth}
    \centering
   \includegraphics[width=\linewidth]{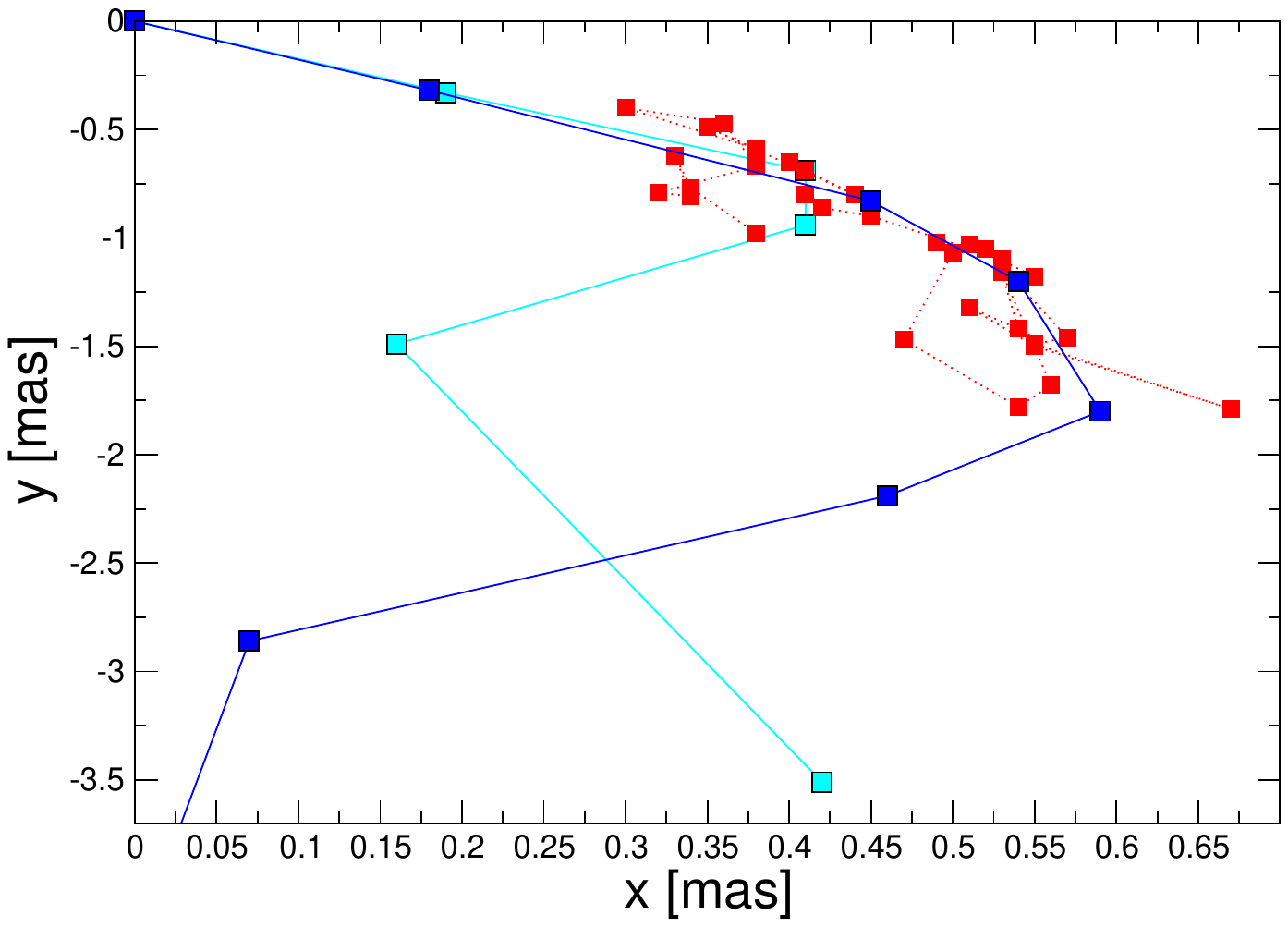}
    [a]
\end{minipage}
\begin{minipage}{\columnwidth}
    \centering
    \includegraphics[width=\linewidth]{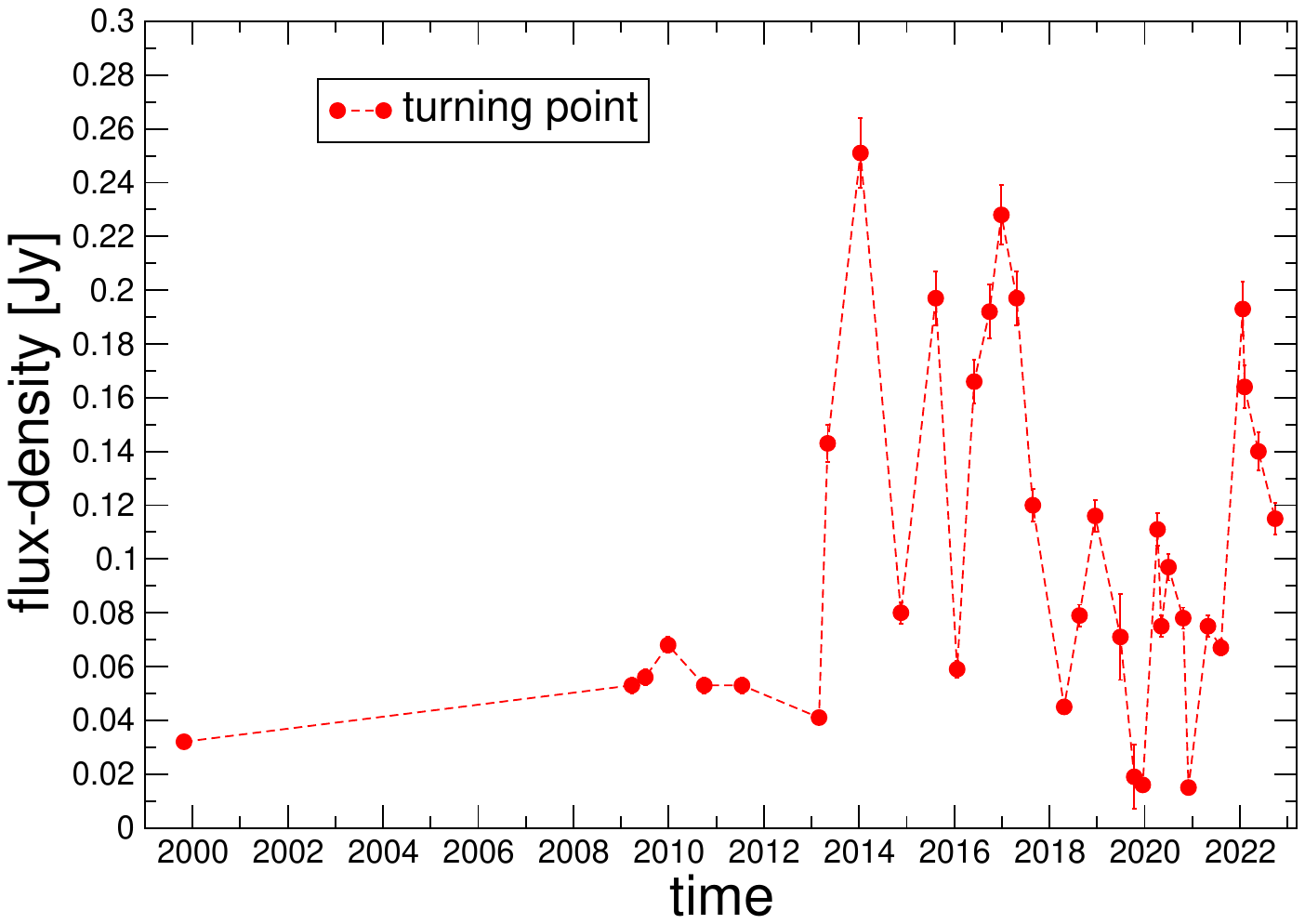}
    [b]
\end{minipage}
 \caption{We visualize the motion of the last jet feature before the jet turns in [a] (red filled squares). Obviously, two turning point regions develop with time. To stress the fact that the region where the jet bends moves with time, we also show the component positions from two epochs of data. The cyan coloured squares (and line) represent the component positions derived for epoch 2016.74, the blue coloured squares (and line) the positions derived for epoch 2022.74. [b] The flux-densities of the jet features at the two turning points in [a] (in red) are shown.}
 \label{encounter}
 \end{figure}
\begin{figure*}
   \centering
 \includegraphics{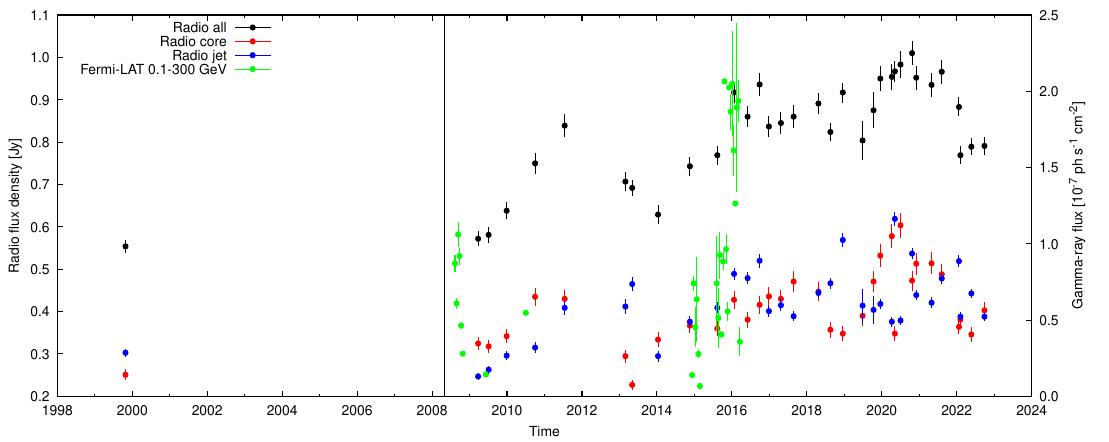}
   \caption{The total flux-density at 15 GHz as obtained from model-fitting the VLBA data (black). In addition, the flux-density of the radio core (red), the sum of the flux-densities of the jet components (blue), and the \textit{Fermi}-LAT $\gamma$-ray flux are shown. The jet contributes similarly as the radio core to the total flux-density. The vertical black line indicates the time of neutrino emission \citep{aartsen}.}%
   \label{total_flux}
    \end{figure*}
\section{Observations, uncertainties \& data analysis}
\label{sec_obs}
\subsection{VLBA data analysis \& Uncertainty estimation}
We re-modeled and re-analyzed 34 VLBA observations (15$\,$GHz, MOJAVE) of PKS\,1717+177 obtained between 1999/12/27 and 2023/05/03.

Gaussian circular components were fitted to all the data to obtain the optimum set of parameters (e.g., flux-density, radial distance, position angle, size of major axis) within the
{\it difmap}-modelfit program \citep{shepherd}. 
Every epoch was fitted independently from all the other epochs.
The model-fitting procedure was performed blindly so as  not to impose any specific outcome. 
Special care was taken to correctly identify the core component in every individual data set. All the component positions derived via model-fitting are displayed in Fig.\,\ref{xy}.

To obtain the uncertainties of the fitted model parameters we employed the approach of calibrating the standard expressions for uncertainties \citep{1999ASPC..180..301F,2008AJ....136..159L,2012A&A...537A..70S} using the tight dependence of the jet component brightness temperature on the component size observed in some sources from the MOJAVE monitoring program (Kravchenko et al., subm.). This is similar to the method of \citet{lister2009} for estimating the positional uncertainty of the model components from the fit of the multi-epoch kinematic data. To account for the uncertainty in the amplitude scale left after a self-calibration we added a 5\% error  to our flux errors in quadrature \citep{kovalev, lister2018}.

\subsection{Optical data analysis}
We obtained optical data for PKS\,1717+177 from the 0.76-m Katzman Automatic Imaging Telescope (KAIT) and from Tuorla Blazar monitoring. 

The KAIT-webpage\footnote{\url{http://herculesii.astro.berkeley.edu/kait/agn/}} provides data and light curves. These light curves are monitored with average cadence of 3 days. The observations are unfiltered and correspond roughly to the R band \citep{li}.

\subsection{X-ray data analysis}
\label{subsec_xray}

PKS\,1717+177 has been observed three times using the \textit{Swift-XRT} telescope. The observations were made in the PC mode in the years 2009 and 2011 with a total exposure time of 13.3\,ks. The combined spectrum was obtained from the \textit{Build Swift-XRT products}\footnote{\url{https://www.swift.ac.uk/user_objects/}} web service \citep{Evans2007, Evans2009} and fitted using \textsc{Xspec} v12.12.0 \citep{Arnaud1996} with an absorbed powerlaw model:
\begin{center}
    \texttt{phabs(zwabs(zpowerlw))}
\end{center}
where the column density of the \texttt{phabs} component was fixed to the Galactic value $n_{\text{H}} = 6.43 \times 10^{20}\;\text{cm}^{-2}$ \citep{Willingale2013}, and redshift of \texttt{zphabs} and \texttt{zpowerlw} components were fixed to $z=0.137$ \citep{SowardsEmmerd2005}. The resulting intrinsic absorption was estimated to $n_{\text{H}} = (7\pm4) \times 10^{20}\;\text{cm}^{-2}$, the photon index to $\Gamma = 1.80^{+0.14}_{-0.13}$, and total unabsorbed flux to $F_{0.2-12\,\text{keV}} = 1.86^{+0.13}_{-0.12} \times 10^{-12}\;\text{erg}\,\text{cm}^{-2}\,\text{s}^{-1}$. We note that the photon index is rather low for a typical BL Lac object \citep{Tavecchio2010, Yang2015}, however, it is consistent within uncertainties with an average photon index of a typical AGN \citep{Gilli2007}. Assuming the luminosity distance of $d_L = 666$ Mpc corresponding to the known redshift ($z = 0.137$) and that the emission is isotropic, the luminosity of the source is $L_{0.2-12\,\text{keV}} = 9.9^{+0.7}_{-0.6} \times 10^{43}$ erg s$^{-1}$.

To describe the feature present in the spectra at $\sim6$ keV (see Fig.\,\ref{fig:xray_spectrum}), we compared the simple absorbed powerlaw model with a model that also includes an emission line. We used a simple redshifted Gaussian line to describe the emission:
\begin{center}
    \texttt{phabs(zwabs(zpowerlw+zgaus))}
\end{center}
where the redshift of the Gaussian line was fixed to $z = 0.137$. The best-fit parameters of the model that includes a Gaussian line were then compared with the previous model in Table \ref{tab:xray}. The intrinsic absorption and photon index are comparable within uncertainties. The centroid of the line was estimated to $6.92^{+0.03}_{-0.05}$ keV. This corresponds to the energy of H-like Fe XXVI indicating the presence of hot plasma, potentially due to shocks. In Fig.~\ref{fig:xray_spectrum} (Appendix~\ref{appendix_xray}), we show the combined, binned X-ray spectrum of the PKS 1717+177 obtained by the Swift-XRT detector with the corresponding spectral fits. 

\renewcommand{\arraystretch}{1.35}
\begin{table*}
    \centering
     \caption{Spectral analysis of \textit{Swift-XRT} data of PKS\,1717+177 that compares two different spectral models: absorbed powerlaw and absorbed powerlaw with Gaussian line.}
     \label{X-ray}
     \begin{tabular}{lllllll}
     \hline
     Model & $n_{\text{H}}$ [$10^{20}$\,cm$^{-2}$] & $\Gamma$ & $F_{0.2-12 \, \text{keV}}$ [erg\,s$^{-1}$\,cm$^{-2}$] & $E$ [keV] & norm [s$^{-1}$\,cm$^{-2}$] & Cstat / dof\\
     \hline
     zwabs(zpowerlw) & $7.1 \pm 4.2$ & $1.80^{+0.14}_{-0.13}$ & $1.86^{+0.13}_{-0.12} \cdot 10^{-12}$ & - & - & 543.2 / 866\\
     zwabs(zpowerlw\,+\,zgaus) & $10.1 \pm 4.2$ & $1.96^{+0.15}_{-0.14}$ & $1.84^{+0.15}_{-0.13} \cdot 10^{-12}$ & $6.92^{+0.03}_{-0.05}$ & $1.5^{+0.7}_{-0.6} \cdot 10^{-5}$ & 529.3 / 864\\  
     \hline
     \end{tabular}
     \label{tab:xray}
 \end{table*}

Apart from the \textit{Swift-XRT} telescope, PKS\,1717+177 was observed once with the \textit{Einstein} telescope and four times within the \textit{XMM-Newton Slew Survey}. Since the flux of the PKS1717+177 is approaching the capabilities of the \textit{XMM-Newton Slew Survey}, only upper limits on the total count rate could be estimated in three cases. 
 Assuming the spectral parameters derived earlier using \textit{Swift-XRT} data, the X-ray fluxes were estimated for all the available observations (\textit{Einstein}, \textit{XMM-Newton Slew}, and \textit{Swift-XRT}) using the HIgh energy LIght-curve GeneraTor (HILIGHT\footnote{\url{http://xmmuls.esac.esa.int/hiligt/}}) in the $0.2-2$ keV, $2-12$ keV, and $0.2-12$ keV energy ranges. The results from these X-ray observations are listed in Table~\ref{X-ray} (see Appendix~\ref{appendix_xray}). 
 
 Due to the relatively poor coverage of PKS\,1717+177 in the X-ray band, it is not possible to study the variability of the source or cross-correlate the X-ray light curve with other energy bands. Nevertheless, based on the available X-ray data, we can see that the X-ray flux of the source remains unchanged within a factor of 1.5 in the years 1980, 2004, 2006, 2009, and 2011. By comparing the data-points to the long-term radio data, we can see that episodes with higher X-ray flux of $F_{0.2-12\,\mathrm{keV}} = (1.71 \pm 0.17) \times 10^{-12}\,\mathrm{erg}\,\mathrm{s}^{-1}\,\mathrm{cm}^{-2}$ (2004, 2009) correspond to the minima in the radio band, while the minimal X-ray flux of $F_{0.2-12\,\mathrm{keV}} = (1.48 \pm 0.11) \times 10^{-12}\,\mathrm{erg}\,\mathrm{s}^{-1}\,\mathrm{cm}^{-2}$ is observed together with maximum flux in the radio band in 2011. The X-ray fluxes in individual epochs are, however, consistent within uncertainties.

\subsection{\textit{Fermi}-LAT data analysis}
In order to obtain light curves of the GeV emission from PKS\,1717+177 (J1719+1745), we make use of observational data from \textit{Fermi}-LAT. We downloaded the photon data from the \textit{Fermi} Science Support Center\footnote{\url{https://fermi.gsfc.nasa.gov/ssc/data/access/}}. Using the script \texttt{make4FGLxml.py}\footnote{Available from the users contributions site \url{https://fermi.gsfc.nasa.gov/ssc/data/analysis/user/}} we generated model files based on the 4FGL catalog version 21. We fit our source of interest with a log-parabola of the form
\begin{equation}
    \frac{\mathrm{d}N}{\mathrm{d}E} = N_0\left(\frac{E}{E_{\mathrm{b}}}\right)^{-(\alpha + \beta\log(E/E_{\mathrm{b}}))}.
\end{equation}
We fixed the break energy\ $E_{\mathrm{b}}$ to its catalog value and left the parameters $N_0$, $\alpha$, and $\beta$ free for the fit. All the parameters of all sources within a radius of $5^{\circ}$ were left free as well. Sources up to a radius of $25^{\circ}$ were included in the model with their parameters fixed to the catalog values. We used the model \texttt{gll\_iem\_v07} for the Galactic diffuse emission and the template \texttt{iso\_P8R3\_SOURCE\_V2\_v1.txt}. This fit was applied to time bins of width of 15~days. Data points with a test statistic (TS) value of 25 or greater\footnote{Corresponding to a detection significance of $\gtrsim 5\sigma$} are plotted as green points in Fig.\,\ref{total_flux}.

  \section{Results}
  \label{sec_results}
 The results of our analysis focus on the detection of the unusual bending traced in the VLBA observations. In Fig.\,\ref{maps}\,[a] and [b] we show two maps of PKS\,1717+177 with Gaussian components superimposed to visualize the general morphology obtained within model-fitting. As obvious from the two images, the jet changed significantly between Fig.\,\ref{maps}\,[a] (epoch 11/01/2014) and Fig.\,\ref{maps}\,[b] (epoch 19/08/2018) and a bending of the jet is observed.
 
 Jet bending is often observed in AGN jets. However, the bending we trace in this particular AGN seems to be caused by an encounter with an otherwise undetected object. This can be seen more easily in Fig.\,\ref{xy_plots}\,[a]-[f]. Here we inform about the evolution of the jet by plotting the jet component positions in xy-coordinates between 1999.82 and 2022.74. Fig.\,\ref{xy_plots}\,[a] shows the earliest data, obtained by model fitting. This first plot shows a jet which starts straight from the radio core (at 0/0) but then bends significantly at a distance of about 0.5 mas (in x-coordinate) from the radio core. Further plots trace the evolution in time by different epochs being displayed in different colours. Obviously the jet appears almost straight up to a core distance of $\sim 0.3-0.5$ mas. From this point onwards, the jet evolves into a zig-zag structure revealing significant evolution with time. 
 
In the following, we describe our results in more detail.

\subsection{VLBA data: A zig-zag jet emerges and develops in PKS\,1717+177}
To understand the phenomenon better, we explore the plots shown in Fig.\,\ref{xy_plots}\,[a]-[f] in more detail.
Plots [a]-[f] show the data for certain periods in time. In 1999.82 ([a]), the jet gets bent. In the following, we call the last jet component of the straight part of the jet the \textit{turning point} of the jet. From this turning point onwards, the jet is bent. This bending leads to a more complex curved jet structure beyond -1.5 mas (y-axis) in the following years (see Fig.\,\ref{xy_plots}\,[b]-[f]). 

The simple bent seen until 2014, evolves into a meandering jet with increasing amplitude of the curvature (plot [f]). While the amplitude of the bending increases, the turning point shifts further away from the radio core. In the latest plot ([f]), the turning point is not a part of a sharp edge (as in plot [c]) any more, but the start of a more gentle curvature at 0.6 mas in the x-coordinate and -1.5 mas in the y-coordinate. 

While the paths of the epochs between 1999.82 and 2021.33 seem to differ, those epochs between 2021.33 and 2022.74 reveal the same path.
The last epoch analyzed (not shown in this paper) indicates a similar bending pattern in 2023.34. We thus trace the emergence and evolution of a zig-zag (or meandering) jet over 23.5 years in time. 

In Appendix~\ref{appendix_jet_paths} (Fig.~\ref{xy}), we show the superposition of the jet components over all the epochs, which can better depict the temporal changes in the position of the turning point and the evolution of the meandering pattern downstream from the turning point.  

In Fig.\,\ref{xy_plots}\,[c], [d], and [e] some features of radio emission are seen on the counter-jet size. These are of low flux-densities and most likely imaging artefacts.

\subsection{Turning point shifts and brightens up in the VLBA data}\label{sec:turnpoint}
To explore the shift of the turning point with time, we perform further studies. In Fig.\,\ref{encounter}\,[a] we plot the xy-positions of those jet components, right before the jet bends (in red). The turning point seems to move in xy-position and two regions can be traced which resemble loops. 

The most recent loop starts at the epoch of 2018.63 and lasts till 2023.34 (the last epoch of the here analyzed data).  By this time, the jet has a similar orientation again, as in 1999.82. Between 1999.82 and 2009, no data are available. Between 2009.23 and 2018.63, we follow how the jet bending evolves and shifts to another bending episode with the turning point being displaced. 

We trace two bending episodes, the first one lasting about nine years (2009-2018), and the second bending episode lasting five years (2018-2023). It is likely, that another, third bending episode, occurred in the time between 1999.82 and 2009, when no data were available. 

To allow for an easier comparison between the turning point and the curved jet structure, we plot in Fig.\,\ref{encounter}\,[a] the component positions from two epochs: 2016.74 (cyan) and 2022.74 (blue).
The turning point reveals strong (possibly periodic) variability as shown in Fig.\,\ref{encounter}\,[b] (in red). We discuss this further in section \ref{sec:variability}.

\subsection{VLBA flux-density evolution}
In Fig.\,\ref{total_flux}, the flux-densities as derived from model-fitting of the 15 GHz VLBA data, are shown (black). In addition, we plot the sum of the jet component flux-densities (blue), the core flux-density (red), as well as the \textit{Fermi} $\gamma$-ray flux (green). The total flux reveals several peaks which are due to an almost equal contribution by the core \textit{and} the jet, as the core is not always the brightest feature.

\subsection{Possible traces for the interaction with an obstacle in the optical light curve}

We detected changes in the optical variability of the object via statistical analysis,  occurring shortly after jet bending.
\subsubsection{Indications in Structure function}
The structure function (SF) serves as a statistical mechanism to assess temporal variability in observed magnitudes of light curves. By gauging the differences between magnitudes over various time intervals, the SF offers a deeper understanding of the light curve's statistical behavior. We hypothesize that if the jet of PKS\,1717+177 comes into contact with an obstacle, influencing either the flux or the flux-density variability, this interaction might manifest as a pronounced change in the SF signal, observable in either the SF characteristic timescales or the amplitudes.

To test this hypothesis, we embarked on a comparative analysis of the SF calculated for the segments of the optical light curve both prior to and during/after the presumed interaction with the obstacle. However, it is pivotal to acknowledge that the accuracy of the SF computation hinges on several parameters, such as the integrity and frequency of the light curve data, along with the persistence and potency of the interaction events.

We  adopted the first-order SF method \citep[see discussion in][and references therein]{kozlowski}, defined as
\begin{equation}
SF(\Delta t)=\sqrt{\frac{1}{N_{\Delta t \mathrm{pairs}}}\sum_{1}^{N_{\Delta t}\mathrm{pairs}}(y(t)-y(t+\Delta t))^{2}},
\label{eq:33}
\end{equation}
\noindent {where   the set of measured magnitudes is denoted by $y=\left\{y_i\right\}, i=1,n$ (e.g., magnitudes) at times $t=\left\{t_i\right\}, i=1,n$ with $\Delta t=|t_{i+1}-t_{i}|$;
and $N_{\Delta t\mathrm{pairs}}$ symbolizes the number of data pairs with the time separation  $\Delta t$.}
We calculated error bars for SF by performing bootstrap resampling of the observed light curve 100 times, and calculating the SF for each sample.
The SFs are calculated for 100 time lag bins across the same time lag range $\sim(0.01, 1000)$ days.

\begin{figure*}
    \centering
\includegraphics[width=0.45\textwidth]{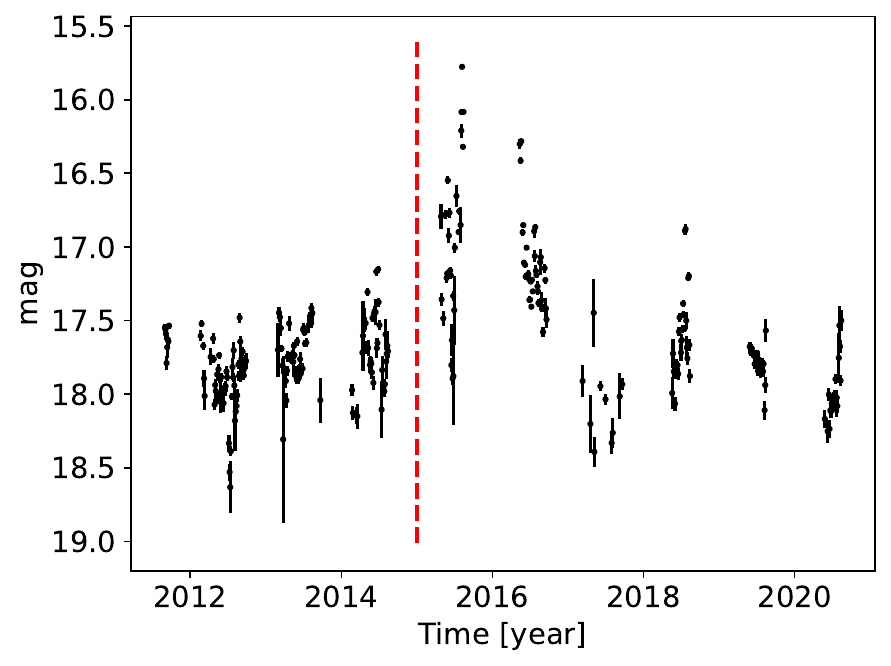}
\includegraphics[width=0.45\textwidth]{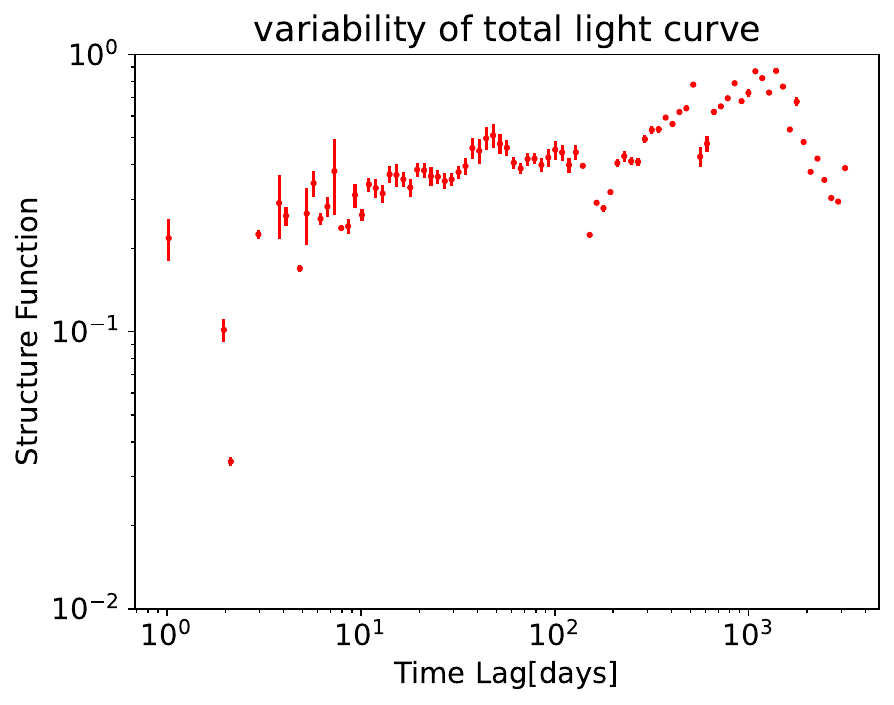}\\
    \centering
\includegraphics[width=0.45\textwidth]{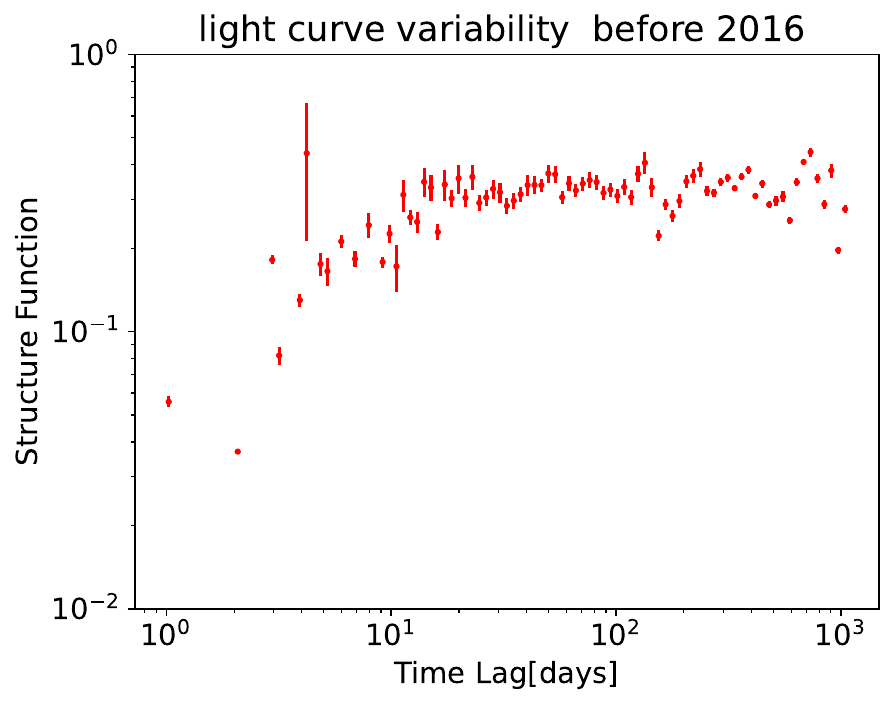}
\includegraphics[width=0.45\textwidth]{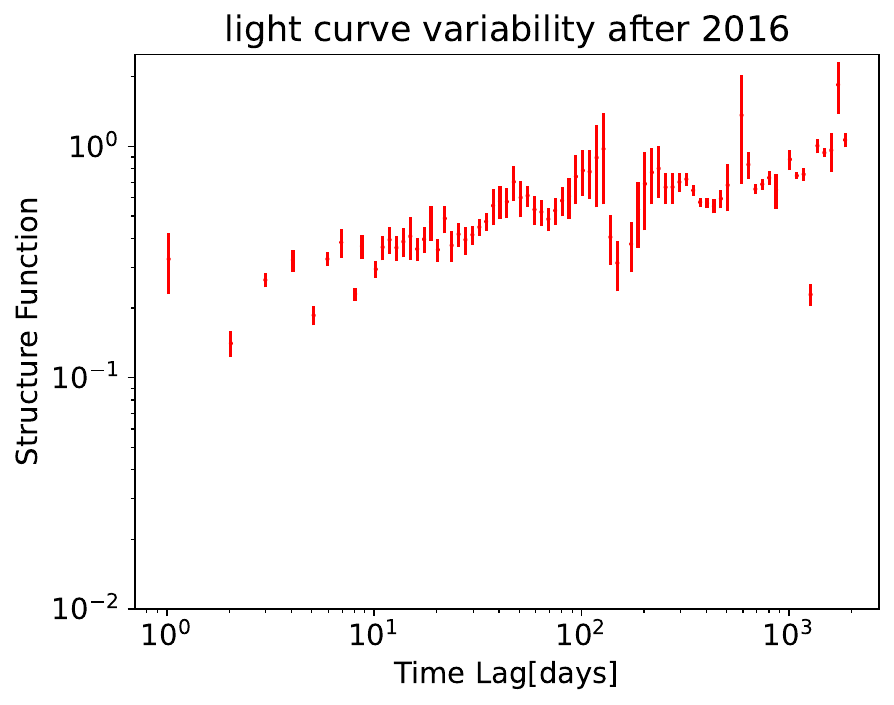}
    \caption{Optical light curve variability. \textit{Upper left}: Optical light curve with errorbars denoting data points. Parts of the light curve associated with weak SF signal (before 2015 yr) and stronger SF  (after 2015 yr)  signal are delineated by the vertical dashed line. \textit{Upper right}: The  SF of the whole optical light curve. The dip in time scales $\sim 100-200$ days corresponds to a part of the light curve around the year 2016.
    \textit{Bottom left}: Partial SF weak signal detected in  the portion of the curve prior to the year 2016. \textit{ Bottom right}: Partial SF stronger  signal detected in  the segment of the curve after 2016.}
    \label{SF_PKS1717}
\end{figure*}

Figure \ref{SF_PKS1717} illustrates the whole optical light curve (upper left plot) separated into sections before  and after 2015, where changes in the SF signal could be expected due to an interaction with an obstacle (see Fig.\,\ref{xy}). The SF of the entire light curve (Figure \ref{SF_PKS1717}, upper right plot) reveals a dip at time scales $\sim 100-200$ days, which corresponds to a part of the light curve around the year 2016. Following this period, the abrupt change in SF values is noticeable, revealing strong SF signal. 
A comparison of the SFs from varying curve segments is illustrated in the lower plots. Given the absence of data for 2015, we opted for 2016 as the demarcation point for computing two sub-SFs. The pre-2016 sub-SF exhibits a subdued signal, whereas the post-2016 segment showcases substantial alterations in SF values, accompanied by increased bootstrap errors.
The larger error bars at shorter time lags, underscores the inherent variability and sampling effects in  quasar light curves \citep[see e.g.,][their Fig.\,5]{decicco}.

 The amplified strength of the post-2016 SF could very well be indicative of an interaction with an external obstacle.
In sum, our analysis of the structure function unveiled prominent discrepancies, pointing towards a shift in the light curve's temporal variability once the jet initiated its deflection. These disparities imply modifications in the observed object's inherent physical processes or behaviors, potentially instigated by the jet's encounter with an obstacle.

\subsubsection{Indications in statistical distributions of magnitudes}
\begin{figure}
    \centering
\includegraphics[width=0.45\textwidth]{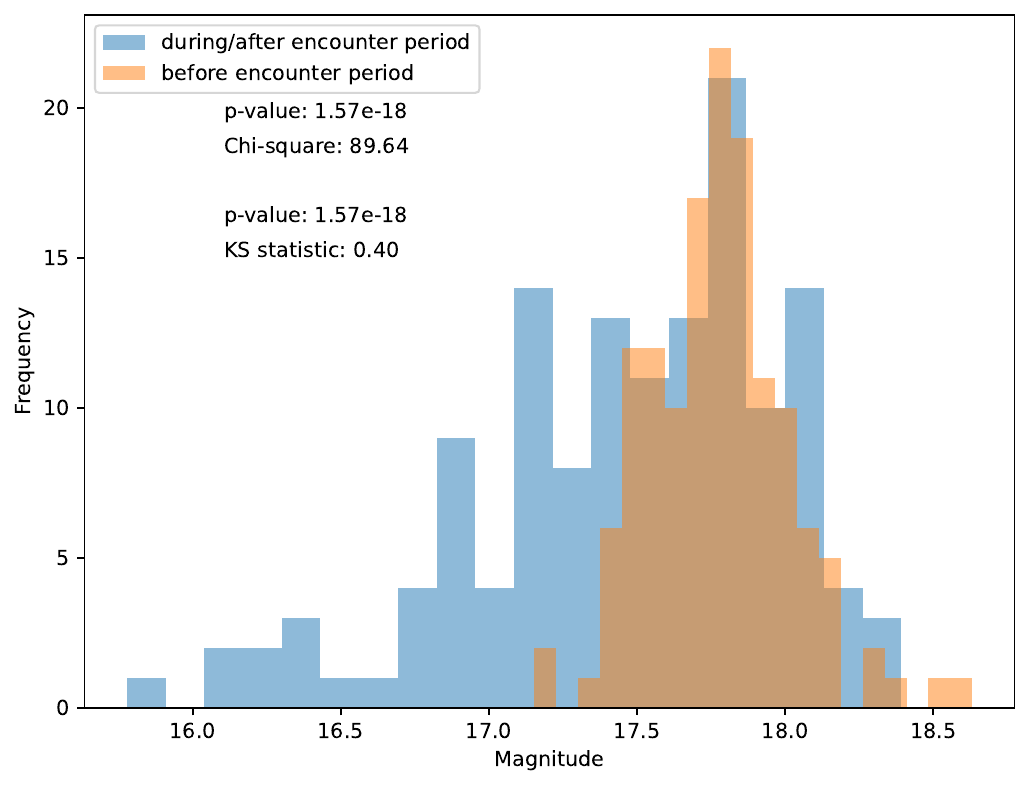}
    \caption{ Histograms of the two categories of magnitudes used for statistical tests: before (orange) and during/after jet encounter with an obstacle (blue). For each histogram, the magnitudes are binned in 20 bins. Calculated test statistics and their p-values are indicated in the plot.}
    \label{test_PKS1717}
\end{figure}

The histogram distributions of magnitudes can reveal variations between the periods before and during/after the jet encounter. To evaluate differences in magnitude distributions across these periods, we employed the $\chi^2$ test and the Kolmogorov-Smirnov (KS) test, segregating magnitudes into pre- and post-encounter groups similarly to the SF analysis (see Fig.\,\ref{test_PKS1717}). The null hypothesis for both tests asserts no significant differences between these distributions. We adopted a 95\% confidence level and used the \texttt{scipy.stats} Python module.

A glance at the histogram for pre-encounter magnitudes (Fig.\,\ref{test_PKS1717}, orange) reveals a narrower distribution, primarily between 17.5-18.5 mag, suggesting consistent variability. In contrast, the post-encounter histogram (Fig.\,\ref{test_PKS1717}, blue) spans 16-18.5 magnitudes, indicating greater variability.

The $\chi^2$ test compared observed magnitude counts in bins to expected counts assuming no association with the jet encounter. The test yielded the  chi-square statistic  of $89.64$, and  $p$-value  $1.57\times 10^{-18}$, indicating a significant discrepancy and a potential association with the jet encounter.

The KS test, assessing the dissimilarity between pre- and post-encounter magnitude distributions, yielded a KS statistic of $0.399$ and a $p$-value of $3.54\times 10^{-10}$. This suggests a significant difference between the distributions, pointing to a possible link between the jet encounter and magnitude variations.

Both tests reveal statistically significant differences in magnitude distributions between the two periods. The low $p$-values indicate that such differences are highly improbable by chance, supporting the theory that the jet's interaction impacts the observed variability in the optical light curve. This is also supported when we apply the methods of the nonlinear analysis to the original and interpolated optical and $\gamma$-ray light curves (see Appendix~\ref{appendix_nonlinear}), in particular the Recurrence Quantification Analysis (RQA). The mean entropy of the interpolated optical and $\gamma$-ray light curves is systematically lower before 2015 than afterwards, see Fig.~\ref{ff2}, which indicates that the system transitions from more ordered to disordered, possible due to the interaction.

\section{Discussion}
\label{sec_discussion}

In this paper, we report the unusual development of the jet of PKS\,1717+177 into a zig-zag structure over 23.5 years in time. 

Jet wiggling is often observed in blazar jets. However, the bending of the jet usually starts with the radio core. A jet, which starts straight and only gets bent at a distance of about 0.5 mas from the core, is atypical and - to our knowledge - has not been observed before on the scales of $\sim 1-10$ parsecs; see, however, the jet bending on the galaxy group scales creating ``dogleg'' quasars \citep{1985ApJ...299..799S}. 

The total radio flux-density variability stems, in almost equal parts, from the radio core as well as from the jet. A significant part of the jet flux-density comes from the almost periodic brightening of the turning point of the jet. Thus, for PKS\,1717+177, the complex jet evolution significantly influences the long-term radio flux-density variability.

\begin{figure}
    \centering
\includegraphics[width=\columnwidth]{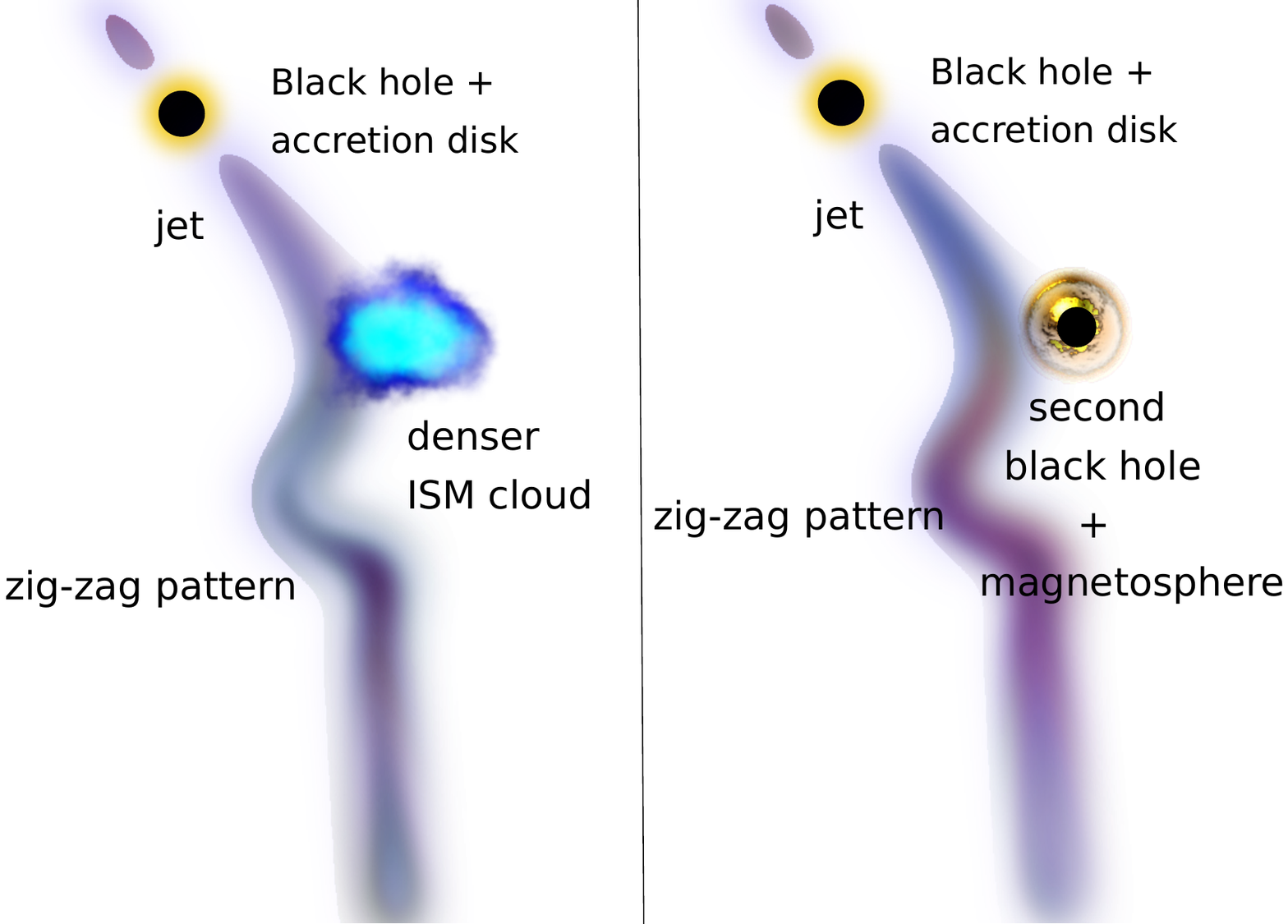}
    \caption{Sketch to illustrate the most likely explanations for the physical nature of the dark deflector positioned at the turning point of the jet.
    An ISM cloud (left panel) or a second black hole with a magnetosphere (right panel) could explain why the straight jets gets bent into a meandering jet structure. This figure is for illustrative purposes only and not to scale.}
    \label{sketch}
\end{figure}

In the following, we discuss a possible origin of this atypical jet bending. 
In particular, we consider, whether Kelvin-Helmholtz instabilities can cause the deflection (section\,\ref{sec:KH}) or if it can be induced by a changing-viewing angle (section\,\ref{viewing}). In section \ref{sec:encounter} we discuss whether a collision with an otherwise undetected object (star, ISM cloud) can explain the pronounced jet curvature. In section \ref{sec:binary}, we explore whether a close pair of supermassive black holes (SMBBH) or the Lense-Thirring effect causes the observed jet bending. In section \ref{sec:second} we discuss the possibility of an interaction with the magnetosphere of a second black hole at the position of the turning point of the jet. For a simple sketch of the two most likely scenarios, please see Fig.\,\ref{sketch}. In section \ref{lensing} we propose the possibility that gravitational lensing by an intervening supermassive black hole causes the meandering of the jet. Finally, in section \ref{neutrinos_process}, we discuss the most likely neutrino generating process.

\subsection{Can Kelvin-Helmholtz or Current Driven instabilities explain the observed jet bending?}
\label{sec:KH}

The observed jet wiggling could be related to instabilities - either Kelvin-Helmholtz (KHI) or Current Driven (CDI), developing in a jet flow and appearing as various helical patterns and oscillations \citep[see][for a review]{2013EPJWC..6102001H}.
For example, depending on the mode, the KHI could disturb the whole jet or mainly its border regions \citep{birkinshaw_1991}. Both types of modes are thought to be observed in the nearby jet in M87 at kpc \citep{2011ApJ...735...61H,2021ApJ...923L...5P} and pc scales \citep{2018ApJ...855..128W,2023Galax..11...33R}, however the helical patterns in the M87 jet were also interpreted as CDI \citep{2017A&A...601A..52B}. Recent high-sensitive VLBA observations at 8 an 15 GHz of the M87 jet \citep{2023MNRAS.526.5949N} and Radioastron observations of the quasar 3C279 at 22 GHz \citep{2023NatAs.tmp....8F} were interpreted as helical threads of the KHI developing in a kinetically dominated flow. The associated disturbances not only amplify the pressure and the magnetic field at the threads location, but could modify the flow velocity making the jet plasma to follow the helical threads \citep{2000ApJ...533..176H}. \cite{2023NatAs.tmp....8F} found that the corresponding Doppler boosting at the specific positions in the helical threads could be observed as jet components at 43 GHz \citep{2017ApJ...846...98J,2022ApJS..260...12W}, while the phase velocity of the disturbance wave corresponds to the observed superluminal motion. Another KHI-related model of the components helical motion includes the disturbance (e.g. shock wave) travelling down the jet threaded by spiral modes of KHI \citep{1995Ap&SS.234...49R}. In this case, the moving shock will lighten up the place of the thread intersection even more, making the appearance of the component travelling down the jet \citep{2012JPhCS.372a2070S}.   

In PKS\,1717$+$177, the apparent jet wiggling starts at the de-projected distance $\approx 1.2/ \sin{\rm \phi}$ milliparsecs from the core, where $\phi$ corresponds to the viewing angle of the jet. Interestingly, this could correspond to the distance where the instability begins to develop. As KHI is suppressed by a longitudinal magnetic field, at some distance from the jet base, the field could drop below the critical value and KHI starts to develop \citep{1995Ap&SS.234...49R}. 

The stability criterion for CDI includes the longitudinal magnetic field that does not depend on the radius, while the jet expansion slows the growth of the instability \citep[][and reference therein]{2013EPJWC..6102001H}. Thus the observed turning point could correspond to the distance where the core of the longitudinal magnetic field forms in MHD models \citep{2023MNRAS.524.4012B,2024MNRAS.528.6046B}. The core formation is expected to operate at 10 -- 100 light cylinder radius (that corresponds to 10$^{2-4}$ $r_{\rm g}$, depending on the black hole spin). 

The expansion profile of the PKS\,1717$+$177 jet, as estimated using both the stacked multi-epoch images \citep[the exponent of the radial width dependence: 1.03$\pm$0.05,][]{2017MNRAS.468.4992P} and size evolution of jet components \citep[the exponent of the components size radial dependence: 1.4$\pm$0.2,][Kravchenko et al. submitted]{lister21} at distances larger than 0.5 mas from the core, is found to be consistent with being conical. According to the present view on the jet acceleration and collimation process \citep{2012rjag.book...81K}, this implies that jet is not magnetically dominated at this scales. Thus it is unlikely that the CDI is responsible for the observed jet wiggling in PKS\,1717$+$178.

The instability model of the PKS\,1717$+$177 jet could also explain why the components seem to avoid some parts of the jet. As suggested by \cite{2023NatAs.tmp....8F} the appearance of the plasma flowing along the helical threads of KHI is affected by the local viewing angle through the Doppler boosting. 

KHI, however, cannot explain the observed quasi-periodic flaring shown in Fig.\,\ref{encounter}\,[b]. In summary, it seems that KHI can explain some, but not all of the observed features.

\subsection{Can viewing angle changes caused by jet bending explain the observed behavior?}
\label{viewing}

PKS\,1717+177 reveals radio flaring at least since 1977, as can be seen in the light curve observed by the University of Michigan Radio Astronomy Observatory\footnote{\url{https://dept.astro.lsa.umich.edu/obs/radiotel/gif/1717_178.gif}} (UMRAO). The flaring pattern we see in Fig.\,\ref{total_flux} seems to have occurred in similar shape earlier already (1992-2008). In case this flaring is caused by precession, we would expect the jet to appear curved already at small distances from the core. However, the jet remains straight up to a core separation of $\sim$0.5 mas. 

Evidence for a changing viewing angle comes from the observed breaks in the components brightness temperature $T_{\rm b}$ dependencies on the core separation $r$ and component size $R$. Kravchenko et al. (subm.) analyzed multiepoch MOJAVE data \citep{lister21} modelled with Gaussian components and found that the exponent of the $T_{\rm b}(r)$ dependence changes from -2.0 to -3.3 at $r = 1.3$ mas. This position corresponds to the typical size of the component at which the dependence $T_{\rm b}(R)$ breaks, changing the exponent from -2.6 to -3.4.

\subsection{Stars and clouds as hydrodynamic deflectors}
\label{sec:encounter}
In addition to the morphological changes seen in the VLBA images, we investigated the optical light curve for modifications of the variability properties which could provide additional support for an interaction of the jet with an unseen obstacle. We show that the SF curve reveals more variability after 2016 compared to the SF curve obtained from observations before 2016. These significant differences in the structure function indicate a change in the temporal variability of the light curve during and after the jet encounter. It seems thus possible that an interaction of the jet with an unseen obstacle like a star, wind, or a cloud changed the physical processes of the jet.

Also, polarization data\footnote{\url{https://www.cv.nrao.edu/MOJAVE/sourcepages/1717+178.shtml}} reveal that there is a strong polarized component travelling down the jet. It appears in the core region near 2011 and can be seen as a decrease of the fractional polarization. Later it can be seen in polarized flux as well. The decrease in the polarization fraction implies that the polarization direction of this component is perpendicular to that of the quiescent jet. This could be an indication of the shock or the disturbance of the magnetic field travelling down the jet. It seems to disappear near 2015, when it approached the jet deflection point $\sim$1.3 mas from the core. We will explore this further in a future work.

There are types of interaction targets that are likely present in most galactic nuclei. These include stars within the dense nuclear star cluster as well as spatially larger structures such as molecular clouds and star clusters \citep[see e.g.][]{2017ARA&A..55...17A}. The nature of the deflection is then purely hydrodynamic, i.e. based on the collision with an obstacle that drives a shock into the jet, which leads to the thermalization of the jet kinetic energy and its slowing down and possible deflection. Subsequently, the jet cone gets broader downstream. These features are qualitatively consistent with the features of the bent jet for the source PKS\,1717+177. Below we briefly discuss two types of targets:
\begin{itemize}
    \item[(i)] individual mass-losing stars, including supernova explosions with extreme velocities,
    \item[(ii)] extended objects (with respect to the jet cross-section), such as molecular clouds and star clusters.
\end{itemize}
Both types of targets (i) and (ii) provide local density and pressure enhancements. The obstacle should be located at the projected distance of $l\sim 1$ mas according to Fig.\,\ref{encounter}\,[a]. For the redshift of $z=0.137$, this corresponds to the projected scale of $\sim 2.44\,{\rm pc}$ (see Table~\ref{tab_source}). Since the source is a blazar, the viewing angle is expected to be small within $\phi\sim 10^{\circ}$, which eventually yields the deprojected distance of $r\sim 2.44\,{\rm pc}/\sin{\phi}\sim 15$ pc, which we consider in the estimates below. The obstacle is expected to be located within the sphere of influence of the SMBH. For the stellar velocity dispersion of $\sigma_{\star}\sim 243\,{\rm km\,s^{-1}}$ for $M_{\bullet}\sim 3\times 10^8\,M_{\odot}$ \citep{2009ApJ...698..198G}, the radius of the sphere of influence is $r_{\rm inf}=GM_{\bullet}/\sigma_{\star}^2\sim 22\,{\rm pc}$.

At this distance, the orbital timescale of the obstacle is, 
\begin{align}
    P_{\rm orb}&=2\pi \frac{r^{3/2}}{\sqrt{GM_{\bullet}}}\,\notag\\
    & \sim 3.14\times 10^5 \left(\frac{r}{15\,{\rm pc}}\right)^{3/2} \left(\frac{M_{\bullet}}{3\times 10^8\,M_{\odot}} \right)^{-1/2}\,{\rm yrs}.\label{eq_orb_period}
\end{align}
assuming that the obstacle is bound to the SMBH within its sphere of influence.
For the orbital velocity of $v_{\rm orb}=(GM_{\bullet}/r)^{1/2}$, the timescale to cross the jet transversely (the distance of $2r\tan \theta$) is,
\begin{equation}
    \tau_{\rm cross}\sim \frac{2r^{3/2}\tan{\theta}}{(GM_{\bullet})^{1/2}}\sim 18\,000 \left(\frac{r}{15\,{\rm pc}}\right)^{3/2}\left(\frac{M_{\bullet}}{3\times 10^8\,M_{\odot}} \right)^{-1/2}\,{\rm yrs}\,,
\end{equation}
hence the set-up with the jet-obstacle interaction is generally stable during the observational monitoring of a few tens of years.

For case (i), the star is parametrized with a mass-loss rate $\dot{m}_{\rm w}$ and a stellar-wind terminal velocity $v_{\rm w}$. When the star is passing across the jet, its relative velocity with respect to the jet plasma flow may be estimated as $\mathbf{v}_{\rm rel}=\mathbf{v}_{\star}-\mathbf{v}_{\rm jet}\approx \mathbf{v}_{\rm jet}$ since the star velocity at 15 pc from the primary SMBH is $v_{\star}=(GM_{\bullet}/r)^{1/2}\sim 300\,{\rm km\,s^{-1}}$, which is much less than the jet plasma speed that is reaching a fraction of the light speed. We scale $v_{\rm rel}$ to $0.1c$. The shock driven into the jet is bow shaped (a bow-shock) whose size is scaled by the stand-off radius $R_{\rm shock}$ from the star \citep[see e.g.][]{1996ApJ...459L..31W,2016A&A...593A.131S,2016MNRAS.455.1257Z,2020ApJ...903..140Z},
\begin{align}
    R_{\rm shock}=\left(\frac{\dot{m}_{\rm w}v_{\rm w}}{4\pi \rho_{\rm jet} v_{\rm rel}^2} \right)^{1/2}\,,\notag\\
    \label{eq_shock_radius}
\end{align}
where $\rho_{\rm jet}=\mu m_{\rm H}n_{\rm jet}$ is the mass density of the jet plasma at a specific distance from the SMBH ($m_{\rm H}$ is the proton mass and $\mu$ is the mean particle weight, here set to 0.5 for an ionized plasma). For hadronic jets, the particle number density is proportional to the jet kinetic luminosity and inversely proportional to the jet plasma speed. The relativistic expression for $n_{\rm jet}$ is as follows,
\begin{equation}
    n_{\rm jet}=\frac{L_{\rm jet}}{\mu m_{\rm H} (\gamma -1)c^2 v_{\rm jet} \pi r^2 \tan^2{\theta}}\,.
    \label{eq_jet_density}
\end{equation}
The jet kinetic luminosity, or rather its lower limit, may be estimated from the radio luminosity at 15 GHz (see Fig.\,\ref{total_flux}). From the flux density of $\sim 0.5$ Jy, we obtain the radio luminosity at 15 GHz of $L_{\rm radio}\sim 3.8 \times 10^{42}\,{\rm erg/s}$. We take this value as an estimate of the jet kinetic power $L_{\rm jet}$, though it is rather its lower limit. The Lorentz factor is denoted as $\gamma=(1-v^2/c^2)^{-1/2}$ in Eq.~\eqref{eq_jet_density}.

First, we consider a star with an intense mass-loss rate and a fast wind speed, e.g. a Wolf-Rayet star with $\dot{m}_{\rm w}\sim 10^{-5}\,{\rm M_{\odot}\,yr^{-1}}$ and $v_{\rm w}\sim 10^3\,{\rm km/s}$, we obtain $R_{\rm shock}\sim 0.007$ pc from Eq.~\eqref{eq_shock_radius}. The ratio between the shock length-scale and the radius of the jet cross-section $R_{\rm jet}\sim r\tan{\theta}\sim 2.6\, (r/15\,{\rm pc})(\theta/10^{\circ})\,{\rm pc}$ then is, $R_{\rm shock}/R_{\rm jet}\sim 0.002$, hence a single star does not result in the planar oblique shock capable of changing the jet direction. 

On the other hand, an extreme mass-loss rate of $\dot{m}_{\rm w}\sim 0.1\,{\rm M_{\odot}\,yr^{-1}}$ and a fast wind of $v_{\rm w}\sim 10^4\,{\rm km/s}$, which are typical for supernova explosions, yield $R_{\rm shock}\sim 2.1$ pc and $R_{\rm shock}/R_{\rm jet}\sim 0.5$. Hence the shock resulting from the supernova explosion and the jet length-scale are comparable in size, which can generally result in the 2D oblique shock capable of changing simultaneously the direction as well as the velocity of the jet flow. The impact of supernova explosions in galactic nuclei were investigated previously, for instance in the context of the impact on the accretion disc \citep{2021ApJ...906...15M}, SMBH feeding \citep{2020A&A...644A..72P}, and the close vicinity of quiescent galactic nuclei, in particular Sgr A* \citep{2015MNRAS.447.3096R}.

In contrast with the single star as a target, with the exception of the supernova explosion, extended obstacles (ISM clouds, star clusters) in case (ii) are more favourable to provide localized density and pressure enhancements on the length-scale of $R_{\rm jet}\sim r\tan{\theta}\sim 2.6\, (r/15\,{\rm pc})(\theta/10^{\circ})\,{\rm pc}$, which is the cross-sectional radius of a jet at the distance $r$ from the SMBH. When the jet moving at the initial Mach speed of $M_{i}=v_{\rm i}/c_{s,i}\gg 1$ interacts with such an obstacle at an oblique angle, the shock driven into the jet leads to the thermalization of its kinetic energy. This is in agreement with the detection of the bright radio component before the point of bending, see Fig.\,\ref{encounter}. 

The detected $H$-like iron line (Fe XXVI) at 6.92 keV could also originate in the shocked plasma at the collision site (see Subsection~\ref{subsec_xray}). The temperature of the post-shock gas can initially reach as much as $T_{\rm shock}\sim 3/16 (\mu m_{\rm H}/k_{\rm B}) v_{\rm rel}^2\sim 10^{10} (v_{\rm rel}/0.1c)^2\,{\rm K}$, which can decrease to $\sim 10^7-10^8\,{\rm K}$ via adiabatic expansion.  

Because of the plasma being thermalized within the shock, the jet slows down significantly -- it is moving at the Mach speed $M_{\rm f}$ with a certain angle of deflection $\alpha$ and within a wide Mach cone with a half-opening angle of $\psi_{\rm Mach}=\tan^{-1}{(M_{\rm f}^{-1})}$. The projected half-opening angle of the cone, in which components move (see Fig.\,\ref{xy}), is about $\psi_{\rm Mach}\sim 38.2^{\circ}$, which results in the post-shock Mach number of the jet flow of $M_{\rm f}\sim 1/\tan{\psi_{\rm Mach}}\sim 1.3$. 

Such an oblique shock then leads to the formation of a bent and widened jet with a disturbed morphology, which was previously denoted as a ``dogleg'' jet \citep{1985ApJ...299..799S}. In comparison with \citet{1985ApJ...299..799S}, who described several ``dogleg'' jets on large, galactic scales (possibly due to the interaction with a halo of another galaxy), the jet of PKS\,1717+177 must clearly interact with an extended object \textit{within} the galactic nucleus, which makes its case unique. Such an obstacle can be a denser clump of $R_{\rm c}\sim 0.1-1.0\,{\rm pc}$ in radius, which has a number density of $n_{\rm c}\sim 10^{5}\,{\rm cm^{-3}}$. 

This is similar to the parameters of clumps within the Circumnuclear Disk in the Galactic center \citep{2021ApJ...913...94H}. At the deprojected distance of $r\sim 2.44\,{\rm pc}/\sin{({10}^{\circ})}\sim  15$ pc, the clump remains within the jet for $\tau_{\rm inter}\sim 2r^{3/2}\tan{\theta}/\sqrt{GM_{\bullet}}\sim 1.8\times 10^4$ years, where $\theta$ is the half-opening angle of the jet (set to 10 degrees) and $M_{\bullet}=3\times 10^8\,M_{\odot}$ is the SMBH mass (see also the estimates below). 

The clump length-scale should be at least $R_{\rm c}\sim r\tan{\theta}\sim 3\,{\rm pc}$ so that an oblique planar shock can form. As the dense clump interacts with the fast and diluted jet plasma, it is continually ablated. The ablation timescale can be estimated following the relation \citep{2012ApJ...750...58B,2022ApJ...931...39M},
\begin{align}
    \tau_{\rm abl}&=\frac{4R_{\rm c}}{q_{\rm abl}v_{\rm rel}}\frac{\mu_{\rm c}n_{\rm c}}{\mu_{\rm jet}n_{\rm jet}}\,\notag\\
    &=2.4\times 10^{10}\left(\frac{R_{\rm c}}{3\,{\rm pc}} \right)\left(\frac{v_{\rm rel}}{0.1c} \right)^{-1}\left(\frac{n_{\rm c}}{10^5\,{\rm cm^{-3}}}\right) \left(\frac{n_{\rm jet}}{1.6\,{\rm cm^{-3}}} \right)\,\text{years}\,,
\end{align}
where $v_{\rm rel}$ is scaled to the jet plasma velocity of $0.1c$ and the jet plasma number density is inferred using the relation given by Eq.~\eqref{eq_jet_density} ($\mu_{\rm c}$ and $\mu_{\rm jet}$ express the mean molecular weights in the molecular cloud and ionized plasma environments, respectively). The ablation coefficient $q_{\rm abl}$ is set to 0.004 following \citet{2011ApJ...739...30K}. We see that $\tau_{\rm abl}>\tau_{\rm inter}$, hence ablation by the jet does not appear efficient enough to destruct the cloud during the passage. 

When the cloudlet is embedded in the hot jet stream with $T_{\rm jet}\sim 10^{10}\,{\rm K}$ set by the inverse Compton limit, it is also subject to evaporation. Typically, the saturation parameter is in this case of the order of unity,
\begin{align}
    \sigma_0=1.84\frac{\lambda_{\rm f}}{R_{\rm c}}\,
\end{align}
i.e. the electron mean free path $\lambda_{\rm f}$ is comparable to the cloud size. In this saturation limit, the mass-loss rate is given by $\dot{m}_{\rm c}\approx 4 \pi R_{\rm c}^2 \rho_{\rm jet} c_{\rm jet}$, where $c_{\rm jet}$ is the sound speed wihin the jet. The evaporation timescale then is,
\begin{align}
    \tau_{\rm evap}&\approx \frac{m_{\rm c}}{\dot{m}_{\rm c}}\,\notag \\
    &\approx 3\times 10^7 \left(\frac{R_{\rm c}}{3\,{\rm pc}} \right)\left(\frac{n_{\rm c}/n_{\rm jet}}{10^5}\right)\left(\frac{T_{\rm jet}}{10^{10}\,{\rm K}} \right)^{-1/2}\,\text{years}\,,
\end{align}
which depends on the exact value of the saturation parameter. However, in principle the evaporating dense cloud can survive long enough within the jet stream.

Due to the velocity shear $v_{\rm shear}$ between the jet plasma and the cloud, Kelvin-Helmholtz (KH) instability develops over time, which can eventually destroy the whole cloud. Given the density ratio between the hot and cold phases $r_{\rho}=n_{\rm jet}/n_{\rm c}\sim 10^{-5}$, the timescale for the development of KH instabilities of the size $\lambda_{\rm KH}\sim R_{\rm c}$ is,
\begin{align}
    \tau_{\rm KH}&=\frac{\lambda_{\rm KH}}{v_{\rm shear}}\frac{1+r_{\rho}}{\sqrt{r_{\rho}}}\,\notag\\
    &\sim 3.1\times 10^4\left(\frac{\lambda_{\rm KH}}{3\,{\rm pc}} \right) \left(\frac{v_{\rm shear}}{0.1c}\right)^{-1}\frac{1+r_{\rho}}{\sqrt{r_{\rho}}}\,\text{years}\,,
\end{align}
which is still longer than a typical crossing time at a given distance. Therefore, the cold dense cloud as a site of the jet collision is stable enough to cause a long-term sharp bending of the jet.

A significant jet deflection due to the interaction with the cloud was also reproduced using hydrodynamic simulations \citep{1999MNRAS.309..273H}. The deflection is usually characteristic of the jets with moderate Mach numbers, small density contrasts with respect to the ambient medium, and smaller kinetic powers. In addition, to sustain a long-term deflection, the ISM cloud should have a sufficient size, i.e. at least exceeding the jet cross-section.

\subsection{Can a close pair of supermassive binary black holes or the Lense-Thirring effect explain the observations?}
\label{sec:binary}
The observations from PKS\,1717+177, particularly the quite regular flaring in the long-term UMRAO light curve, and the bent jet (Fig.\,\ref{xy_plots}), offer a dynamic framework of core astrophysical processes. Also, the observed unusual equivalency in radio emission between the jet and the radio core (see Fig.\,\ref{total_flux}) may be indicative of a binary black hole system’s influence, e.g., through dynamical interactions leading to jet precession (e.g., \citet{britzen_2023} and references therein), increased particle acceleration due to shocks, and potential dual jet contributions, further influenced by modulations due to orbital motion and interactions with surrounding accretion material.

We outline two compelling scenarios: the influence of a secondary black hole in a binary system (SMBBH), and the effects surrounding a single SMBH with a misaligned accretion disk.

In Fig.\,\ref{encounter}\,[b]), we observe a distinctive brightening at the jet’s turning point, that could be associated with periodic variations in the accretion rate, leading to alternating phases of enhanced and diminished jet luminosity. The brightening at the turning point could result from increased accretion during the close approach of components, where tidal forces pull more material into the accretion disk of the component, intensifying the jet emission.

The long-term variability seen in the radio regime (UMRAO) can also be attributed to the binary nature of the SMBBH. As the black holes orbit, the changing distance between them alters the gravitational influence on the surrounding accretion disk. During closer approaches, enhanced gravitational tidal forces can increase the accretion rate, leading to a surge in emitted radiation, while at farther distances, the accretion rate and consequently the emission diminish. This oscillation in gravitational influence correlates with the observed periodic variability in the jet's flux, offering a possible link between the binary SMBH dynamics and the jet’s luminosity variations.

For example, considering a primary black hole mass of $10^9 M_{\odot}$, the binary hypothesis would yield a mutual separation on the order of 132 Schwarzschild radii.
Such a separation could be commensurate with, or even less than, the expected dimensions of the primary's accretion disk, raising the possibility of disk truncation or significant perturbations due to tidal forces exerted by the secondary black hole.

The observation of a `sharp edge' followed by a smooth deflection in the jet’s turning point may result from an instantaneous increase in the accretion rate, induced by the heightened tidal forces during the close approach of the binary components. This increased accretion rate intensifies jet activity, producing a sharp increase in emission. As the black holes move apart in their orbit, the tidal forces reduce, leading to a decrease in the accretion rate and a subsequent smooth deflection as the jet activity stabilizes.


An alternative scenario, posits a single SMBH with a misaligned accretion disk. 
Here, Lense-Thirring precession \citep{1975ApJ...195L..65B} - a relativistic effect wherein the inner part of the accretion disk experiences precession due to the spin of the black hole - may influence the jet. The resultant bent jet could be an outcome of the jet's attempt to reorient itself with the spin axis of the black hole, particularly if there is a significant misalignment between the accretion disk and the black hole's spin. However, this explanation does not straightforwardly account for the $\gamma$-ray light curve delay observed in Section~\ref{appendix_nonlinear}, necessitating  additional processes.

While the binary SMBH hypothesis offers an intuitive and cohesive explanation for the observed features of PKS\,1717+177, yet the single SMBH scenario remains a feasible, although unconventional, option. 
\subsubsection{Why does the jet start straight and only gets bent at larger core distance?}
 
 Gravitational fields from components in the binary can generate a perturbation region between the Roche lobes, notably near the L1 Lagrange point, where both black holes exert important gravitational effects. Here, tidal forces may influence the jet, causing it to bend due to differential gravitational interactions. Furthermore, shifts in the mutual positions of the binary components can vary the gravitational field the jet encounters, resulting in its dynamic bending over time. Given that our object is at a redshift of 
0.137 ($\sim 500$ Mpc), and if we assume the total mass of the system to be $10^{8}M_{\odot}$, with period of 4.5 yr, a physical shift due to motion would be 0.02 light days in the rest frame. However, observed shifts of 0.05-0.1 mas, under assumption of period of 4.5 yr would yield an unreasonably high total mass ($10^{10}-10^{11} M_{\odot}$
). 
  This could also explain why the bending starts slightly away from the core, where the secondary black hole's influence might become more prominent.
 Also, the orientation of the binary's orbital plane relative to the orientation of the jet could play a role in where the bending initiates. If the orbital plane is misaligned with the jet's direction, this misalignment might result in the observed bending pattern. Additionally, precession of the binary's orbital plane over time could introduce variations in the jet's bending behavior.

The partially straight jet can also be explained alternatively in the scenario of the misaligned disk.
The jet might interact progressively with the gravitational forces of the misaligned disk. As the jet moves outwards, the net effect of these interactions might result in the observed bending pattern at some distance from the core. Internal jet processes, like acceleration or turbulence, could initially counteract deflection, maintaining the jet's straightness near the core until the misaligned disk's effects manifest.

Another possibility mirrors the binary SMBH scenario, where a massive star or even dark matter clump orbits a single SMBH. Their gravitational influence might create localized regions that can deflect the jet's trajectory. The jet might display a specific bending pattern based on the gravitational interactions along its path. The star or dark matter's periodic orbit around the SMBH could cause repeated gravitational interactions with the jet, leading to the possibility of observing even the periodicity in its bending. However, this scenario is complex and necessitates further multi-wavelength  observations.

Support for the here sketched scenario of a possible supermassive binary black hole comes from \citet{oneill}. The authors find evidence for a supermassive binary black hole (SMBHB) candidate in the blazar PKS\,2131-021, based on periodic flux-density variations. They suggest an SMBHB with an orbital separation of $\sim$0.001–0.01 pc. Interestingly, their 15 GHz VLBA maps (MOJAVE) also indicate a straight jet on parsec-scale. 
\subsection{Could the jet bending be caused by a second black hole at the turning point of the jet?}
\label{sec:second}
In galaxy evolution, mergers of galaxies are expected. The last stage of the galaxy merger is the bound pair of two massive black holes. The second black hole can approach the primary one from any angle, and thus the interaction between the jet launched by the primary SMBH cannot be excluded. Once the second black hole becomes bound and starts orbiting the primary one at a random orbital inclination, the chance that it is on the orbit that intersects the jet with a half-opening angle $\theta$ is as follows,
\begin{equation}
    f_{\rm int}\sim \frac{2/3 \pi r^3 \tan^2{\theta} (2\pi/4\theta)}{4/3 \pi r^3}\,,
    \label{eq_jet_BH_int}
\end{equation}
where the numerator takes into account the volume taken up by the two jet cones along the whole orbit of the second black hole, hence the term $2\pi/(4\theta)$ takes into account that the black hole will encounter the jet on its orbit (not only at a specific epoch). The denominator considers the whole volume of the region at the distance $r$ of the second black hole from the primary one. From Eq.~\eqref{eq_jet_BH_int} it follows that it does not depend on the distance, i.e. $f_{\rm int}\sim 1/2 \tan^2{\theta}(2\pi/4\theta)=\pi\tan^2 \theta/(4 \theta)\sim 0.14$ or in 14\% of SMBH-SMBH mergers in jetted AGN, such an interaction is expected (for the half-opening angle of 10 degrees) under the assumption that the jet direction is approximately constant during the orbital period of the order of $10^5$ years, see Eq.~\eqref{eq_orb_period}.

A naive hypothesis is that the jet bending is caused by the gravitational field of the second massive black hole. In this case, the plasma is flowing around the second black hole at a fraction of the light speed, $v_{\rm jet}\sim 0.1c$. Therefore, the gravitational effect of the second black hole may be estimated via the Bondi-Hoyle-Lyttleton lengthscale as follows,
\begin{equation}
    r_{\rm BHL}\sim \frac{2GM'_{\bullet}}{v_{\rm jet}^2}\sim 9.6 \times 10^{-5} \left(\frac{M'_{\bullet}}{10^7\,M_{\odot}} \right)\left(\frac{v_{\rm jet}}{0.1c} \right)^{-2}\,{\rm pc}\,,
\end{equation}
where we scaled the second black-hole mass $M_{\bullet}'$ to $10^7\,M_{\odot}$. The scale $r_{\rm BHL}$ is a small fraction of $r_{\rm BHL}/(r\tan{\theta})\sim 3.6 \times 10^{-5}$ of the total transversal dimension of the jet body, hence the gravitational effect is negligible for realistic black-hole masses. Therefore, the only way to significantly deflect the jet is via the magnetosphere of the second black hole.

\subsection{Testing the binary scenario with magnetic field-jet interaction} 
\label{sec:magnetic}
\begin{figure}
    \centering
    \includegraphics[width=0.45\textwidth]{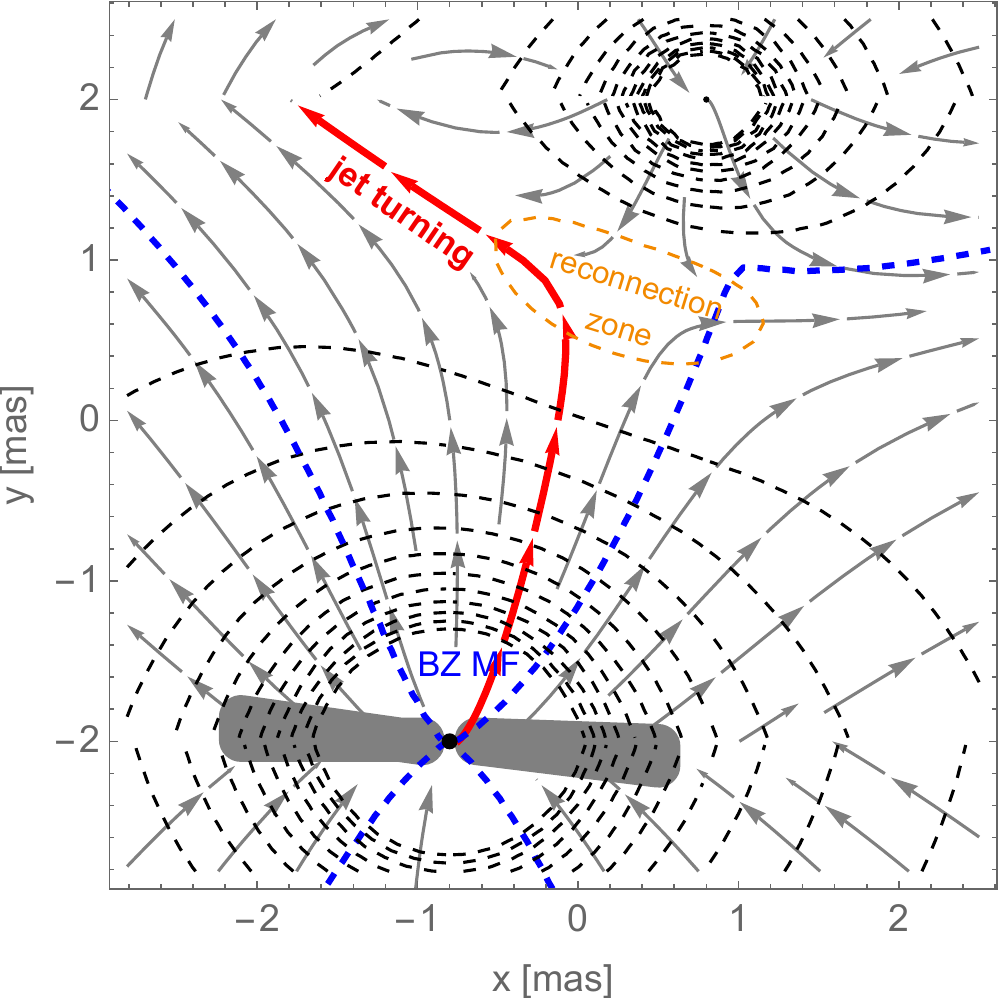}
    \caption{Jet--second black-hole magnetosphere interaction scenario. A primary heavier black hole of $5\times 10^8 M_\odot$ with accretion disk (black dot and gray polygon at the bottom of the figure) generates a large-scale vertical magnetic field of a parabolic shape (gray arrows) interacting with the magnetic field of the second black hole of the mass $5\times10^6 M_\odot$. The blue dashed curves indicate the boundary magnetic field lines entering the edge of the black hole ergosphere, separating the jet-launching region driven by the Blandford-Znajek (BZ) mechanism. The red arrows indicate a possible jet trajectory along the magnetic field lines, which is turning when approaching the second lighter black hole (on the top of the figure). The black dashed lines indicate the equipotential contours of the magnetic field strengths. The orange dashed contours show the magnetic reconnection zone due to the existence of the magnetic null points. }
    \label{fig:binaryMF}
\end{figure}

The observed sharp turning of the jet can also be caused by the interaction of the jet with the external magnetic field component of a lighter companion black hole. 
The topology of magnetic field plays a significant role in jet launching and propagation mechanisms. The black hole jets, which are usually modeled by the Blandford-Znajek (BZ) mechanism \citep{1977MNRAS.179..433B}, require the presence of a large-scale vertical magnetic field, which can arise from magnetorotational dynamo processes, driving away the energy and angular momentum through plasma outflow \citep{2023arXiv231100034J}.  

If the system is a supermassive binary, one can expect each of the components to form its own magnetosphere if the separation distance is large enough. Since efficient jet launching requires specific magnetic topology, it is not necessary that both SMBBH components have jets, especially for large mass ratios of the black holes. However, the presence of a strong magnetic field source at large distances from the primary jet launching black hole will inevitably cause a change of the magnetic field topology along the directions towards the second black hole. The resulting field will be given by superposition of the corresponding components of the two magnetic field sources. Similarly, it will affect the jet propagation from the primary SMBH, causing it to turn, especially if the jet is initially launched towards the second black hole. 

To model the black hole magnetosphere we consider simple parabolic solution often utilized in numerical GRMHD and PIC simulations and other jetted black hole setups \citep[see, e.g.][]{Tch-Nar-McK:2010:APJ:,Cri-etal:2020:AA:,2023EPJC...83..323K}
\begin{equation} \label{A-parab}
    A_{\phi} = \frac{B}{2} r^{k} (1- |\cos\theta|), 
\end{equation}
where $A_{\phi}$ is a vector potential, $B$ is the strength of the magnetic field, $k\in[0,1.25]$ is declination of the field lines. Depending on the choice of $k$, one can restore the BZ parabolic model for $k=1$ and the BZ
split-monopole for $k=0$ \citep{1977MNRAS.179..433B}. To describe the jet launching region, $k=3/4$ is widely used. In Fig.\,\ref{fig:binaryMF} we plot a possible scenario of the jet turning in the region, where the large-scale magnetic field of a primary black hole is connected with that of the second black hole. Both magnetic field components are modeled using the potential (\ref{A-parab}), with the resulting global magnetic field governed by a supporposition of two solutions, i.e. $A_{\phi (1)}+A_{\phi (2)}$. The strengths and radial dependencies of the two field components are scaled relative to the corresponding black hole masses, which are chosen to be $M_1 = 5\times 10^8 M_{\odot}$ for the primary jet-launching black hole, and $M_2 = 5\times 10^6 M_{\odot}$ for the second black hole. The global shape of the magnetic field lines depends on the ratio of the strengths of the two field components, which is chosen to be equal to the ratio of masses of the two black holes. 

A precise jet trajectory, as well as the location of the turning point, and turning angle of the jet, depends on the initial choice of the magnetic field configuration of the two black hole components, their relative directions, the black hole mass ratio, and the initial pitch angle of the jet with respect to the magnetic field of the second black hole component. Therefore, in Fig.\,\ref{fig:binaryMF} we demonstrate one of the many possible realizations of the described scenario. For larger mass ratios of the black hole components, the turning point has to be closer to the second (lighter) black hole. This is because the black hole's magnetic field is expected to decay more rapidly with the actual distance for lighter black holes. Therefore, even if the magnetic field strength at the event horizon scale of the lighter black hole is stronger than that of the primary black hole, it will decay faster with distance than that of the primary one. On the other hand, if the mass ratio is extreme (e.g. $\gg 100$), the cross-section of interaction of magnetic field components can be too small to cause the turning of the jet. Therefore, scenarios with a lighter but still supermassive black hole companion are more plausible for the observed pattern than those with stellar mass objects.

If the second black hole is located not too far from the turning point of the jet, which is at a distance of $0.4 - 1.5$ mas from the primary black hole, the orbital shift of the second black hole cannot be greater than $10^{-2}$ mas per year for the values given in Table~\ref{list}. This implies that the relative position of the binary would change only slightly during the 23.5 years of observations. This also implies that the jet launched towards the second black hole would always start straight. However, since the turning angle of the jet in a suggested model is highly sensitive to the pitch angle of the jet and the magnetic field direction of the second black hole, whose dynamical timescale is expected to be much shorter than that of the primary one, it is highly probable that the further propagation of the jet after turning point will be different in different epochs, as we indeed observe. 

Another interesting implication of suggested scenarios is that the combination of different magnetic fields of the binary components may lead to the appearance of magnetic null points or regions of a minimal magnetic potential if the two field components are not aligned in the same direction. Magnetic null points play a crucial role in magnetic reconnection theory and related particle acceleration scenarios. The region of a minimal magnetic potential, where the magnetic null points can exist, is marked by an orange dashed loop in Fig.\,\ref{fig:binaryMF}. The jet encountering matter in this region is expected to produce a high-energy emission in the form of neutrino and $\gamma$-rays.

\subsection{Loops and regular outbursts at the turning point}
\label{sec:variability}

\begin{figure*}
    \centering
    \includegraphics[width=0.4\textwidth]{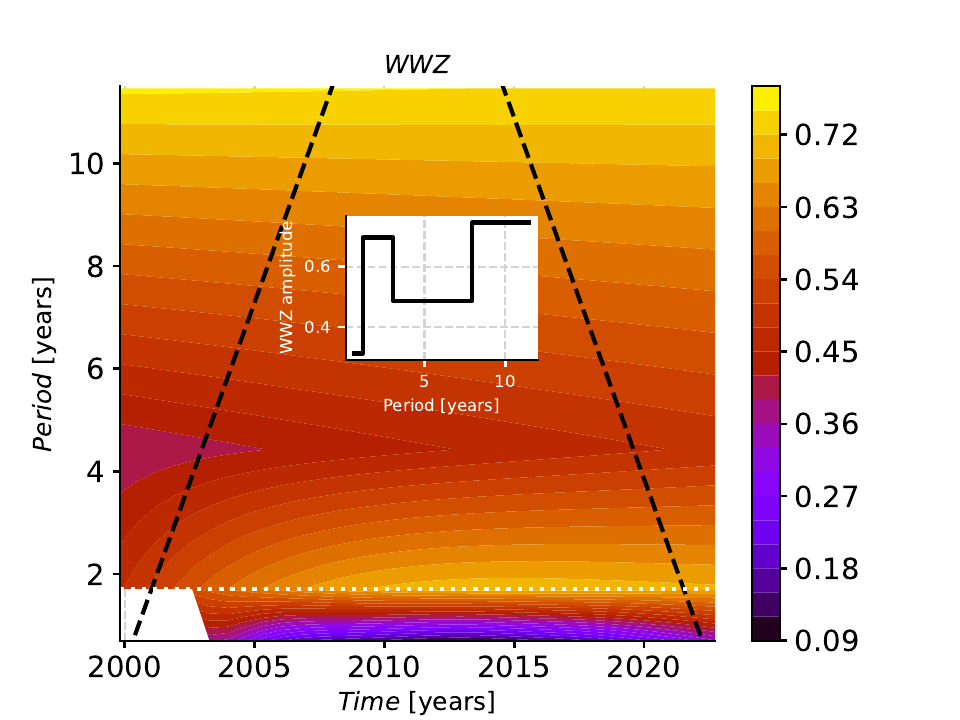}
    \includegraphics[width=0.5\textwidth]{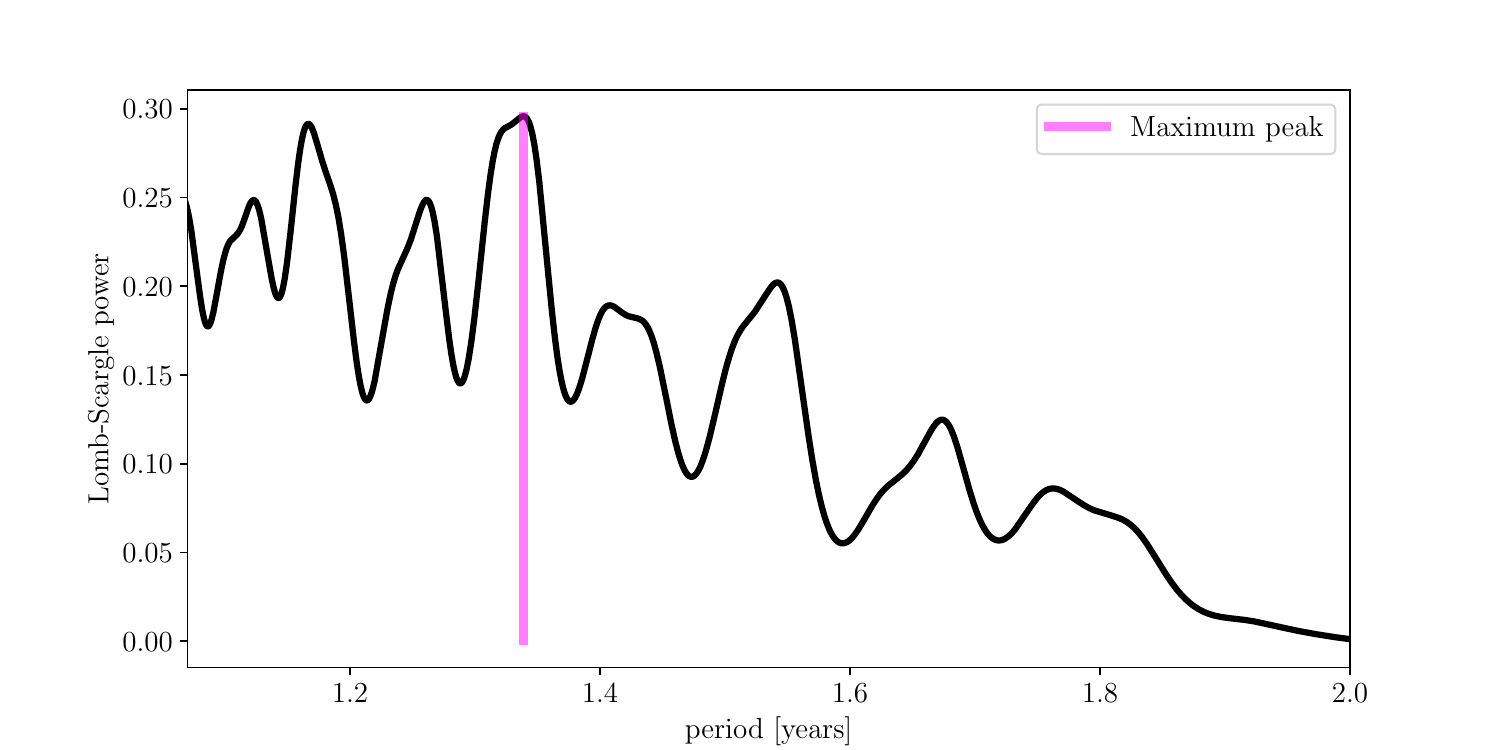}
    \caption{Indication of the periodicity of the last component before the jet turns. \textit{Left panel:} The weighted wavelet $z$-transform indicates the peak in the WWZ amplitude in the period-time plane (both expressed in years). This peak is at the period of $\sim 1.71$ years (dotted white line). The dashed black lines indicate the cone of influence. \textit{Right panel:} Lomb-Scargle periodogram with the most prominent peak at $1.34$ years.}
    \label{fig_period_interaction}
\end{figure*}

The projected distance of the turning point changes over the timescale of $(2022.74-2016.74)\sim 6$ years, see Fig.~\ref{encounter}. The last component before the jet bend also performs a smaller, oscillatory-like motion  (see Fig.~\ref{encounter}, top panel), which is manifested by the loop-like motion with the projected loop size of $\sim 0.089 \times 0.649$ mas. This oscillatory-like motion is related to regular flux outburts with the characteristic timescale of $\sim 1.3-1.7$ years, see Fig.~\ref{encounter} (bottom panel). The timescale close to $1.5$ years is also revealed using the weighted wavelet $z$-transform method (WWZ, see Fig.~\ref{fig_period_interaction}, left panel) and the Lomb-Scargle periodogram (Fig.~\ref{fig_period_interaction}, right panel).

The changes and variability could be interpreted as the manifestations of the interaction of the jet with the dark obstacle, in particular the ISM cloud, which leads to the oblique shock. Then the last component before the jet bend would correspond to the emission of the shock due to the thermalization of the kinetic energy of the jet.
The shift in the distance from the radio core could be related to the cloud inhomogeneities, i.e. clumpiness. In particular, the shift over the six years implies the clump size of $l_{\rm clump}\sim v_{\rm orb}\tau=\sqrt{GM_{\bullet}/r}\tau\sim 1.8\,(M_{\bullet}/3\times 10^8\,M_{\odot})^{1/2} (r/15\,{\rm pc})^{-1/2} (\tau/6\,\text{years})$ milliparsecs or $\sim 370$ AU. When the jet with a given kinetic power encounters a denser cloud, the shock forms closer to the SMBH because of the increased thermal pressure. When the jet-clump interaction stops due to either the orbital motion or the clump evaportation, the jet penetrates deeper into the ISM cloud and shock forms at larger distances.

The loop-like motion with the projected radius of $R_{\rm loop}\sim 0.325\,{\rm mas}\sim 0.8\,{\rm pc}$ could be interpreted by the rotation of the jet -- the shocked gas is pulled by the toroidal magnetic field within the jet sheath. The jet rotation in that case would have an angular velocity of $\omega= 2\pi/P_{\rm rot}\sim 240^{\circ}\,{\rm yr^{-1}} (P_{\rm rot}/1.5\,\text{years})$. The regular outbursts would then correspond to the Doppler-boosted emission by a factor of $\lesssim 5$ of the rotating component according to the relation $S=S_0\delta(\gamma, \phi)^{p+\alpha}$, where $S_0$ is the intrinsic component emission, $\delta$ is the Doppler-boosting factor depending on the Lorentz factor and the viewing angle, $p$ expresses the component geometry ($p=3$ for a spherical component), and $\alpha$ is the power-law spectral index \citep[see e.g.][]{britzen_2023}. Another possibility is that the loop-like motion is the manifestation of the jet wiggling due to propagating Alfvén waves along the jet. Using the expression for the Alfvén velocity, $v_{\rm A}=B/(4 \pi \mu n_{\rm jet}m_{\rm H})^{1/2}$, where $n_{\rm jet}$ is the number density of a hadronic jet according to Eq.~\eqref{eq_jet_density}. Using the estimate for the radius of the oscillation, $R_{\rm loop}\sim 0.8\,{\rm pc}$, and the expression for the Alfvén timescale, $\tau_{\rm A}\sim R_{\rm loop}/v_{\rm A}\sim 1.5$ years, we obtain the required magnetic field strength of $B\sim R_{\rm loop}\sqrt{4\pi \mu n_{\rm jet}m_{\rm H}}/\tau_{\rm A}\sim 0.2\,{\rm G}$ in order for the Alfvén waves to induce a sufficiently rapid jet bending, which is a moderate value at the scale of $\sim 15$ pc from the radio core. More generally, the loop-like pattern can trace accretion-disc/jet instabilities as well as the perturbations propagating from the radio core, for example caused by an orbiting secondary component on smaller scales, see Subsection~\ref{sec:binary}.




\subsection{Can gravitational lensing by an intervening black hole explain the observations?}
\label{lensing}





In the previous sections we discussed the possibility that the jet is locallly - within the source - deflected. We also want to discuss another possibility which involves gravitational lensing by an intervening object. The observed meandering radio emission may result from a combination of intrinsic jet emission and a modified jet structure, influenced by gravitational lensing due to a second but distant - 
black hole. This lensing hypothesis is supported by the two loop-like structures seen in the motion of the turning point in Fig.\,\ref{encounter}\,[a] and the rapid flux-density flaring in Fig.\,\ref{encounter}\,[b]. Moreover, the detection of absorption lines in the spectrum of PKS 1717+177, an otherwise featureless BL Lac object, suggests the presence of an intervening absorber along the line of sight \citep{SowardsEmmerd2005}. Assuming the dark deflector is a second black hole and that we observe a BL Lac object at a small angle to the line of sight, gravitational lensing is expected.

The loop-like structures with a length-scale of $\sim 0.6\,{\rm mas}$ at the longer side can be associated with the lensing effect of the second black hole. If the distance to the source is $D_{\rm J}\sim 650.7\,\rm Mpc$ and the length-scale between the lens--black hole and the jet is within $D_{\rm LJ}\sim r\sim 1\,{\rm kpc}$, from the Einstein radius formula, we obtain the constraint on the lens mass of,
\begin{align}
    M_{\bullet}'&=\frac{c^2\theta^2}{4G}\frac{D_{\rm J}^2}{D_{\rm LJ}}\,\notag\\
    &\sim 5\times 10^{9} \left(\frac{2\, \theta}{0.6\,{\rm mas}} \right)^2 \left(\frac{D_{\rm J}}{650.7\,{\rm Mpc}} \right)^2 \left(\frac{D_{\rm LJ}}{1\,{\rm kpc}} \right)^{-1}\,M_{\odot}\,. 
    \label{eq_mass_lens}
\end{align} 
The less massive black hole of $\sim 10^8\,M_{\odot}$ would need to be located on the galactic scales of $\sim 40 {\rm kpc}$ to cause a similar lensing effect. A lens of $\sim 5\times 10^6\,M_{\odot}$ (supermassive black hole, globular cluster, giant molecular cloud) would need to be located at $\sim 1 {\rm Mpc}$ to provide the required lensing. 

The second massive black hole as a deflector  can address the intriguing pattern of the last jet component before the bending, see Fig.\,\ref{encounter}\,[a]. This loop-like motion could also be associated with quasi-regular pattern in the radio light curve corresponding to the last component, see Fig.\,\ref{encounter}\,[b], which exhibits the repetition timescale of $P_{\rm int}\sim 1.3-1.7$ years. This can be shown from both the weighted wavelet $z$-transform (see Fig.\,\ref{fig_period_interaction}, left panel) and the Lomb-Scargle periodogram  (see Fig.\,\ref{fig_period_interaction}, right panel), which both indicate the peak close to $\sim 1.3-1.7$ years. The radio flaring could be associated with the shock that is orbiting around the intervening massive black hole. Each time the shock emission is moving towards the observer, its emission is Doppler-boosted.


The presence of an intervening black hole may also affect the detection of neutrinos from PKS 1717+177, as gravitational lensing can enhance the number of neutrino events. A quantitative estimate of this enhancement could be made if the magnification factor of the lens were known.  However, this value remains uncertain. We leave a detailed study of the potential enhancement of neutrino events due to gravitational lensing in the PKS 1717+177 source for future work.

\subsection{What is the most likely neutrino generating process?}
\label{neutrinos_process}
The motivation to study this blazar in more detail originated from the coincidence in position with a high-energy neutrino detected by IceCube \citep{aartsen} and the hope to find the deciding peculiarity which hints at the neutrino generating process.

The most atypical and astounding peculiarity in this blazar is the jet-bending by an unseen deflector. We here briefly discuss, whether this deflection might be connected with the neutrino generation in this source and develop ideas within both scenarios, cloud and SMBBH. 
Unfortunately, the detected neutrino arrived before the radio data were monitored regularly (Fig.\,\ref{total_flux}).

Two scenarios might be connected with the generation of neutrinos. Either the cloud-scenario, or the SMBBH-scenario.

A collision scenario might lead to an enhanced production of neutrinos in the cloud-model. The shock within the cloud can accelerate particles via turbulence or magnetic reconnection at the jet/cloud interface. At the same time, the shock-generated X-ray photons could interact with the surrounding hadrons and in this p-$\gamma$-process generate neutrinos. However, this requires a high X-ray energy density to provide a sufficient number of energetic photons to increase the likelihood of generating neutrinos \citep[see e.g.][]{2024PhRvD.109j1306M}. 

The high-ionization H-like line of Fe XXVI at 6.92 keV is present in the X-ray spectrum, which indicates the origin in a strong shock that can accelerate protons to high enough energies. The 0.2-12 keV X-ray luminosity of $\sim 10^{44}\,{\rm erg\,s^{-1}}$ is also substantial. Hence, the basic conditions for the neutrino generation via the p$\gamma$ mechanism are met.

Within the SMBBH-scenario we propose that the superposition of both black hole magnetospheres creates a reconnection zone, where neutrinos can be generated via the interaction of accelerated protons and the energetic photons that can also arise in shock-heated plasma as in the previous scenario.

 Hence, in both scenarios a neutrino can be generated via the p$\gamma$ process. The final assessment of the neutrino-generation mechanism depends on the observational confirmation of the presence of the extended cloud intersecting the jet path or the second SMBH whose magnetosphere bends the magnetized jet plasma.

\section{Conclusions}
\label{sec_conclusions}

We summarize the main findings about the blazar PKS\,1717+177 below:
\begin{itemize}
\item PKS\,1717+177 reveals a rather slim pencil beam, until a significant bending of the original jet path sets in. The temporal evolution of this bending can be traced over 23.5 years. After 23.5 years, the jet returns to its original simple bend, which was also seen in 1999.82. However, we cannot be sure, to not have missed another episode - the complex bending evolution might happen more often and with a shorter period. 
\item The variability seen in the optical light curve (by means of a SF analysis) changes around 2016, with the variability becoming stronger after 2016. 
\item During the complex bending evolution, the turning point shifts. Almost periodical flaring of the radio emission (at 15 GHz), as derived from the VLBA data, is observed at this turning point of the jet. 
\item We discuss several scenarios to explain the observed unique combination of properties: the pronounced and complex bending, the loop-like motion of the turning point, the quasi-regular flaring at the turning point, and the change in the variability properties of the optical light curve.
\item Several scenarios can explain part of the observed phenomena but not all of them (e.g., KH-instabilities; a supermassive binary black hole within the radio core; the LT-effect; an encounter with an unseen and dark object like a supernova or an ISM cloud). A dense circumnuclear cloud as a deflector might provide a likely explanation due to the properties of the ISM cloud and commonness of infalling ISM clouds in galactic centers, as deduced from observations of the Galactic Center in the Milky way (e.g., \citet{zadeh}).
\item Another scenario involves a jet--second black-hole magnetosphere interaction that can explain the bending, as well as the $\gamma$-ray flaring and neutrino production.
\item A magnetohydrodynamic deflection of the jet by an ISM cloud or a second massive black hole at the turning point can explain the intriguing pattern of the last jet component before the bending: the loop-like motion (Fig.\,\ref{encounter}\,[a]) as well as the quasi-regular pattern in the radio light curve (Fig.\,\ref{encounter}\,[b]). The radio flaring could then be associated with a helical motion of the shock due to the jet rotation. As the shock emission is quasi-periodically enhanced due to Doppler boosting, the turning point exhibits the variability pattern with the typical timescale of $\sim 1.3-1.7$ years.
\item The observed meandering radio emission and loop-like structures in the jet, supported by absorption lines in the spectrum, are consistent with gravitational lensing by an intervening supermassive black hole. The lensing effect could also enhance the number of neutrino events, although a precise quantitative estimate depends on the unknown magnification factor. Further studies are needed to fully explore the role of gravitational lensing on both the radio and neutrino observations.
\item Neutrino production in PKS 1717+177 is plausibly associated with the jet colliding with the dark deflector (an ISM cloud or a second black-hole magnetosphere), which results in a strong shock and an associated large X-ray emission energy density. The shock site can naturally create conditions for the interaction of the X-ray photon field with the protons accelerated due to turbulence or magnetic reconnection.
\end{itemize}
While black hole pairs on large scales have been studied across the wavelength regimes e.g., \cite{komossa, krause19}, direct imaging of close pairs with high resolution radio interferometry remains elusive. To our knowledge, only one pair with 7.3 pc separation has so far been imaged \citep{rodriguez06}. In case the dark deflector in PKS\,1717+177 is a supermassive black hole, then we have detected a pair of supermassive black holes with one radio-loud black hole (radio core with jet) and a dark black hole (deflector). This would be a unique supermassive binary system, with an even smaller separation of less than 2 parsecs (this is a lower limit due to jet bending).

It is also worth performing additional multiwavelength observations of PKS 1717+177 to search for more signs of the shocks due to the interaction of a relativistic jet with a dense and spatially extended molecular cloud, such as high-ionization lines in the X-ray spectrum or forbidden line transitions. In particular, the temporal evolution of the spectrally resolved X-ray Fe XXVI line could confirm its origin at the jet-cloud interaction site and could provide hints about the density and the homogeneity of the deflector.

\section*{Acknowledgements}
We thank the anonymous referee and the internal reviewer S. v. Fellenberg for constructive comments which improved the manuscript.
The authors are thankful for very helpful and inspiring discussions
with A. Eckart, V. Karas, P. Biermann, and A. Witzel. MZ acknowledges the support from the GA\v{C}R-LA grant No. GF23-04053L. E.K. and A.T. thank the Alexander von Humboldt Foundation for its Fellowship. This work has made use of public \textit{Fermi} data obtained from the High Energy Astrophysics Science Archive Research Center (HEASARC), provided by NASA Goddard Space Flight Center. Tuorla Blazar Monitoring Program (HTTP://users.utu.fi/kani/1m) is supported by Academy of Finland projects 317636 and 320045. This research has also made use of data from the MOJAVE database, which is maintained by the MOJAVE team \citep{lister}. The National Radio Astronomy Observatory is a facility of the National Science Foundation operated under cooperative agreement by Associated Universities, Inc. RP would like to express his acknowledgement for the institutional support of the Research Centre for Theoretical Physics and Institute of Physics, Silesian University in Opava, SGS/30/2023, and for the support from the project of the Czech Science Foundation GA{\v C}R \mbox{23-07043S}. TP is supported by GA{\v C}R grant 21-13491X. L.\v C.P. and A.B.K. are supported by the Ministry of Science, Technological Development, and Innovation of R. Serbia through projects of Astronomical Observatory Belgrade (contract 451-03-47/2023-01/200002) and the University of Belgrade - Faculty of Mathematics (contract 451-03-68/2023-14/200104).
FJ acknowledges funding by the Austrian Science Fund (FWF) with project SOFT [P35920]. IP is supported in the framework of the State project ``Science'' by the Ministry of Science and Higher Education of the Russian Federation under the contract 075-15-2024-541.

\section*{Data Availability}
The data underlying this article will be shared on reasonable request to the corresponding author.







\appendix
\section{Superposition of the observed jet paths}
\label{appendix_jet_paths}
\begin{figure*}
   \centering
 \includegraphics[width=15cm]{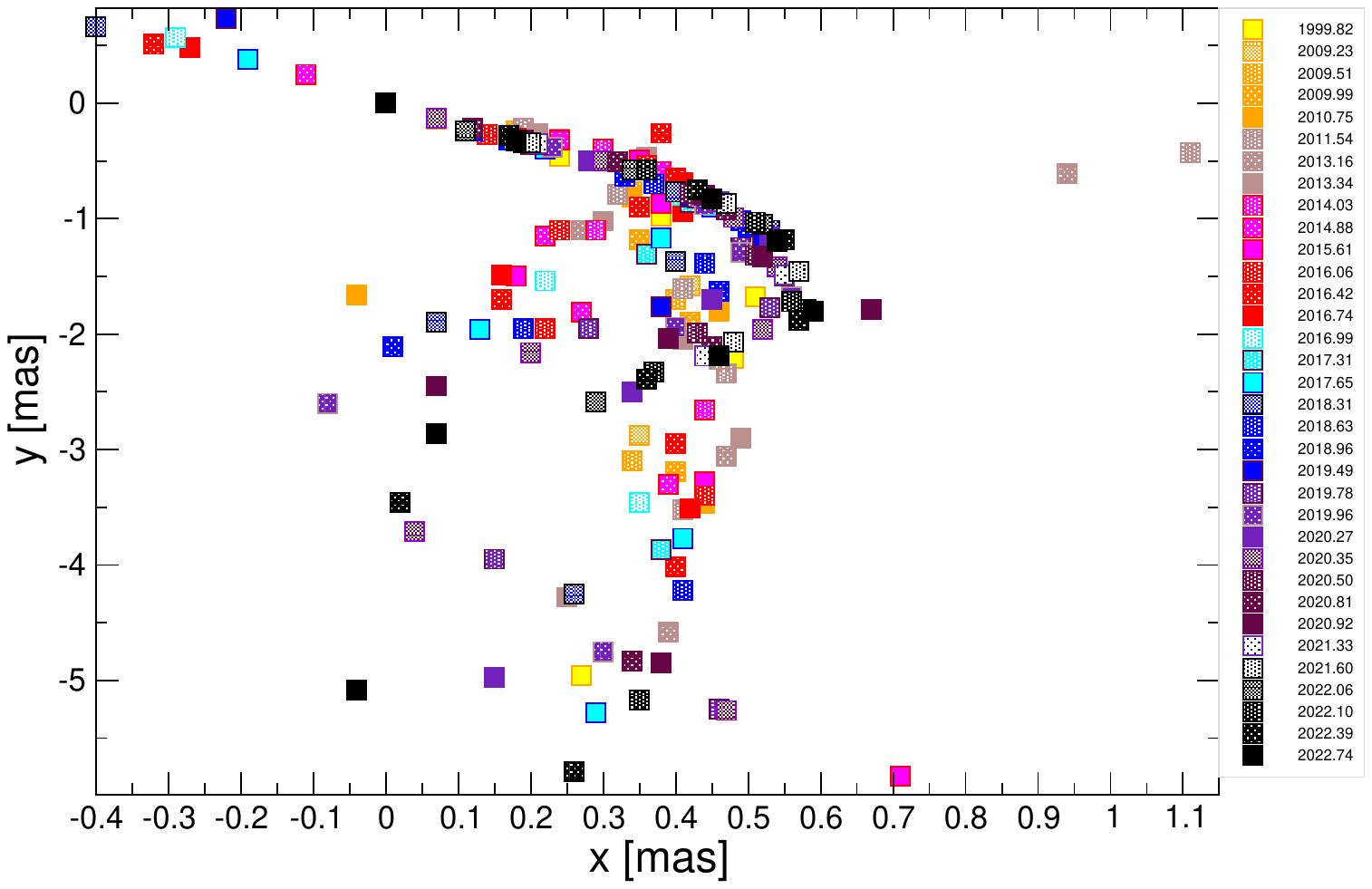}
   \caption{All xy-coordinates of jet component positions from Fig.\,\ref{xy_plots}\,[a]-[f] superimposed. This plot indicates that the bending evolved with time and that the deflection point started at a larger core distance, shifted towards a smaller distance with time and then moved to larger core distances again. Please note the different scales on the x- and y-axis. }%
   \label{xy}
    \end{figure*}

Because the zig-zag jet evolves (larger amplitude) and the turning point shifts in position with time, the superposition of all jet components in Fig.\,\ref{xy} wrongly appears at first sight as a disrupted jet. Only a more detailed analysis of the temporal jet evolution can provide insight into the physical mechanism at work which leads to a continuous transformation of the jet.
\section{X-ray observations of PKS\,1717+177.}
\label{appendix_xray}

\begin{figure*}
    \centering
    \includegraphics[width=0.85\textwidth]{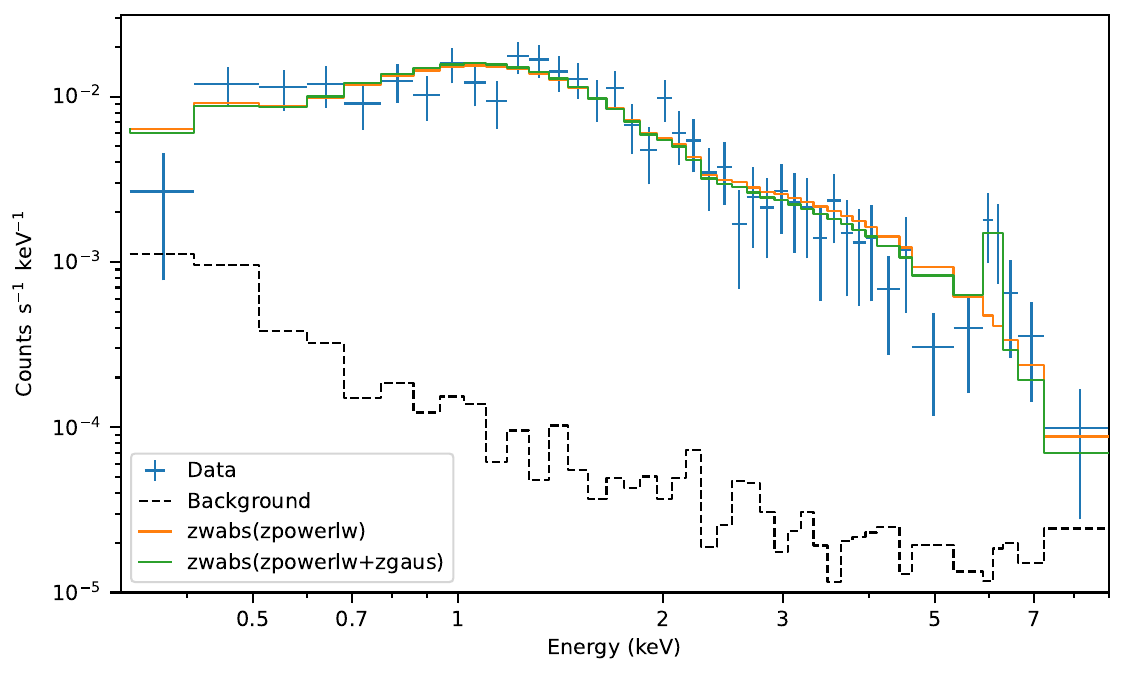}
    \caption{Combined X-ray source spectrum of PKS\,1717+177 (blue) and the background spectrum (black), as observed by the \textit{Swift-XRT} telescope. The spectrum was fitted using a simple absorbed powerlaw model (orange) and absorbed powerlaw with a Gaussian line (green). The spectrum was binned for visualization purposes.}
    \label{fig:xray_spectrum}
\end{figure*}

In Fig.~\ref{fig:xray_spectrum}, we present the combined X-ray spectrum of PKS 1717+177 obtained by the Swift-XRT telescope. In Table~\ref{X-ray}, we list the X-ray observations performed by the Einstein, XMM-Newton, and Swift-XRT telescopes. 

\begin{table*}[h!]
    \centering
     \caption{X-ray observations of PKS\,1717+177. The observation dates are given in the format year/month/day and HH:MM:SS UTC.}
     \label{X-ray}
     \begin{tabular}{llllll}
     \hline
     \hline
     Detector & Observation Date & Exp. time [s] & Energy range [keV] & Count rate [$\mathrm{s}^{-1}$] & Flux [$10^{-12}\,\mathrm{erg}\,\mathrm{s}^{-1}\,\mathrm{cm}^{-2}$]\\
     \hline
     Einstein & 1980/09/24 & 7706 & $0.2-2$ & $0.031 \pm 0.003$ & $0.584 \pm 0.005$\\
    \hline
      XMM-Newton & 2004/09/05  10:00:37 & 7.57 & $0.2-2$ & $<0.80$ & $<1.06$\\
         &  &  & $2-12$ & $<0.49$ & $<4.39$\\
        &  &  & $0.2-12$ & $0.63 \pm 0.31$ & $1.83 \pm 0.91$\\
     \hline
      XMM-Newton & 2006/09/17  07:59:27 & 3.22 & $0.2-2$ & $<1.26$ & $<1.67$\\
         &  &  & $2-12$ & $<0.64$ & $<5.74$\\
        &  &  & $0.2-12$ & $<1.65$ & $<4.81$\\
     \hline
        XMM-Newton & 2011/09/13  07:07:09 & 3.63 & $0.2-2$ & $<0.82$ & $<1.08$\\
        &  &  & $2-12$ & $<0.51$ & $<4.59$\\
        &  &  & $0.2-12$ & $<0.97$ & $<2.83$\\
     \hline
        XMM-Newton & 2012/09/28  04:58:36 & 1.74 & $0.2-2$ & $<0.78$ & $<1.03$\\
         &  &  & $2-12$ & $<1.11$ & $<9.89$\\
         &  &  & $0.2-12$ & $<0.95$ & $<2.78$\\
     \hline
        Swift-XRT & 2009/01/08  12:36:11 & 4809 & $0.2-2$ & $0.028 \pm 0.003$ & $0.65 \pm 0.07$\\
         &  &  & $2-12$ & $0.009 \pm 0.002$ & $0.89 \pm 0.17$\\
         &  &  & $0.2-12$ & $0.037 \pm 0.003$ & $1.70 \pm 0.15$\\
     \hline
        Swift-XRT & 2011/08/17  01:16:04 & 4044 & $0.2-2$ & $0.025 \pm 0.003$ & $0.58 \pm 0.07$\\
         &  &  & $2-12$ & $0.010 \pm 0.002$ & $0.94 \pm 0.18$\\
         &  &  & $0.2-12$ & $0.034 \pm 0.003$ & $1.56 \pm 0.15$\\
     \hline
        Swift-XRT & 2011/08/19  04:20:13 & 3954 & $0.2-2$ & $0.020 \pm 0.003$ & $0.46 \pm 0.06$\\
         &  &  & $2-12$ & $0.011 \pm 0.002$ & $1.07 \pm 0.20$\\
        &  &  & $0.2-12$ & $0.031 \pm 0.003$ & $1.40 \pm 0.15$\\
     \hline
     \hline
     \end{tabular}
 \end{table*}
 
\section{Nonlinear Analysis of optical, and $\gamma$-ray data}
\label{appendix_nonlinear}
Nonlinear time series analysis of observational data is used to gain more insight into the dynamical properties of PKS\,1717+177. Limited data, however, challenges the nonlinear analysis of unique objects like blazars. 

We address the impact of missing data by comparing two approaches: using the raw data without addressing missing values (NaNs), and on the other hand, applying linear interpolation to fill the missing values.  Interpolation is employed to estimate the values of missing data points based on the surrounding observed data, allowing for a complete dataset, which is more suitable as input for nonlinear methods of analysis. While interpolation introduces artificial dynamics into the data, comparing measures derived from interpolated data with those obtained from raw data  offers more insights into the dynamics, potentially enhancing their reliability, particularly when they display similar values.  To handle outliers and manage missing values to achieve a uniformly distributed light curve, we utilized the AstroPy and SciPy library in Python  \citep{2013A&A...558A..33A, 2020SciPy-NMeth}. The results obtained from these methods are labeled with a lowercase "E" index in Tab. \ref{tab} and Figs.  \ref{ff3} and \ref{ff2}.

Recurrence Quantification Analysis (RQA) is a technique within nonlinear time series analysis used to characterize recurrent patterns' structure \citep{2007PhR...438..237M}. It provides insights into the dynamical properties and underlying mechanisms of physical systems. RQA quantifies the recurrence of states in phase space, which refers to the tendency of a system to revisit similar states over time. By analyzing the properties of recurrent patterns, such as their density, duration, and distribution, RQA can reveal information about the system's behavior. It has applications in various fields, including physics, biology, finance, and engineering, where understanding the dynamics of complex systems is crucial.
In simple terms, it is a numerical representation of the recurrence plot \citep{1987EL......4..973E}, which is a graphical tool for studying  properties of phase space trajectories of dynamical systems.

The fundamental RQA measures we employ here are: Recurrence Rate (RR),  which quantifies the concentration of recurrent points within a phase space.
Determinism (DET), which is the ratio of recurrence points forming diagonal lines to the total recurrence points, where a higher DET indicates a deterministic system. Entropy (ENTR) calculates the Shannon entropy of the probability distribution of diagonal line lengths. It assesses the complexity or uncertainty of the system. A higher ENTR value signifies a more intricate or less orderly system, while a lower value indicates greater orderliness.

In this work, we utilize a modified version of RQA \citep{2023EPJST.232...47P}, with the goal of providing more accurate measurements of noisy and irregular data. Such a novel application of RQA was used in \citet{bhatta} and \citet{britzen_1502}, where RQA measures are considered as a function of the RR and averaged.  The RR is related to the threshold value that determines which two points in phase space are denoted as recurrent. Instead of choosing a single fixed threshold value to calculate the RQA measures, this approach considers information from 50 threshold values corresponding to particular RRs, which can be seen as various levels of recurrences. More specifically, the value of DET and ENTR was calculated for every RR $ \in [1,2,\dots, 50]$  and then averaged to capture the dynamics of the system's behavior on multiple scales, including higher levels.  Therefore the RQA measures of determinism and entropy are denoted as mDET and mENTR, where "m" stands for mean, see \citep{2023EPJST.232...47P} for more details.

In the analysis of time series data embeddings, following Takens theorem \citep{10.1007/BFb0091924, 2015Chaos..25i7610B}, which suggests that the phase space of a nonlinear system can be reconstructed from a one dimensional time series by embedding it into a higher-dimensional space. We estimated the optimal time lag (tau) using the average mutual information (AMI) method \citep{Kantz2005NonlinearTS}.  Subsequently, we determined the embedding dimension (emb) using the L.Cao algorithm \citep{1997PhyD..110...43C}.

\subsubsection{Results of the Non-linear Analysis}

\begin{figure*} 
		\centering 
		\includegraphics[width=0.9\linewidth]{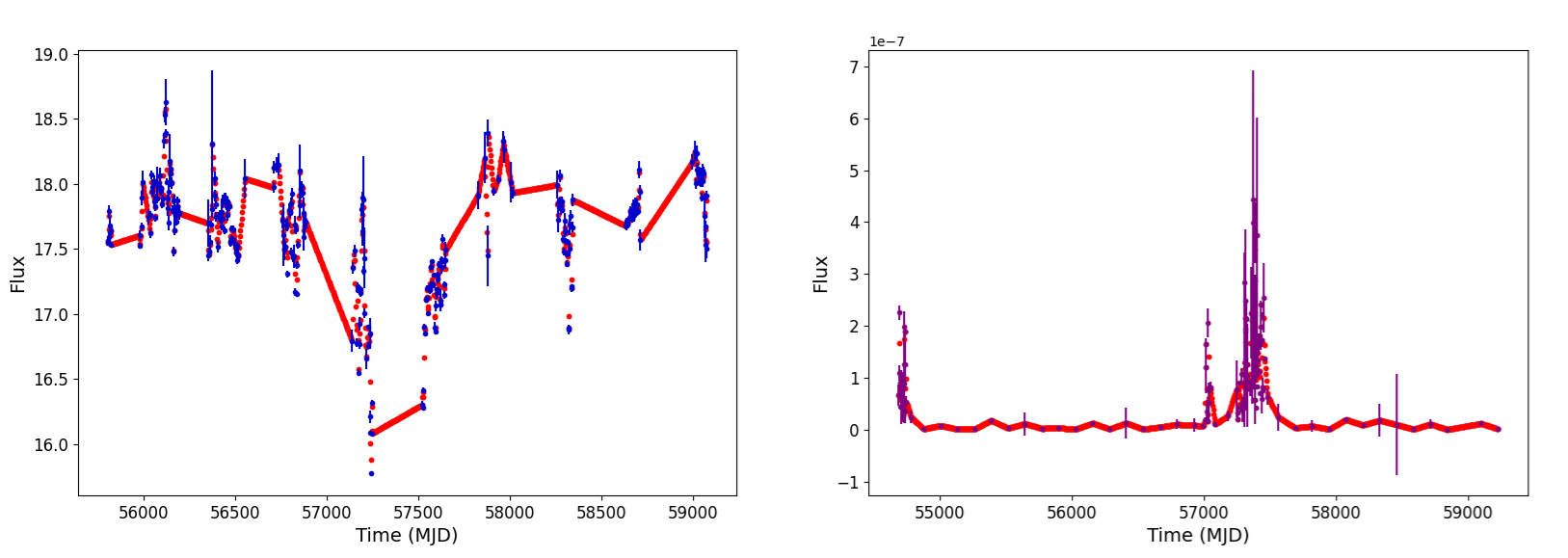}

\caption{
This plot shows the analysed light curves (optical (left): blue, $\gamma$-rays (right): purple). The red points indicate the flux values that were added to achieve evenly binned light curves. The sampling bins were determined based on the median values of the bin distributions, resulting in values of 3.008 and 4 for the optical and $\gamma$-ray data, respectively. The percentage of the artificially added points in Tab.\,\ref{tab} in  "inter" column. Upon aligning the evenly sampled light curves to the same time frame with common binning, minimal correlation was observed between emissions at higher and lower energy levels.
}		 \label{ff1}
\end{figure*}

\begin{figure*}
		\centering 
		\includegraphics[width=1\linewidth]{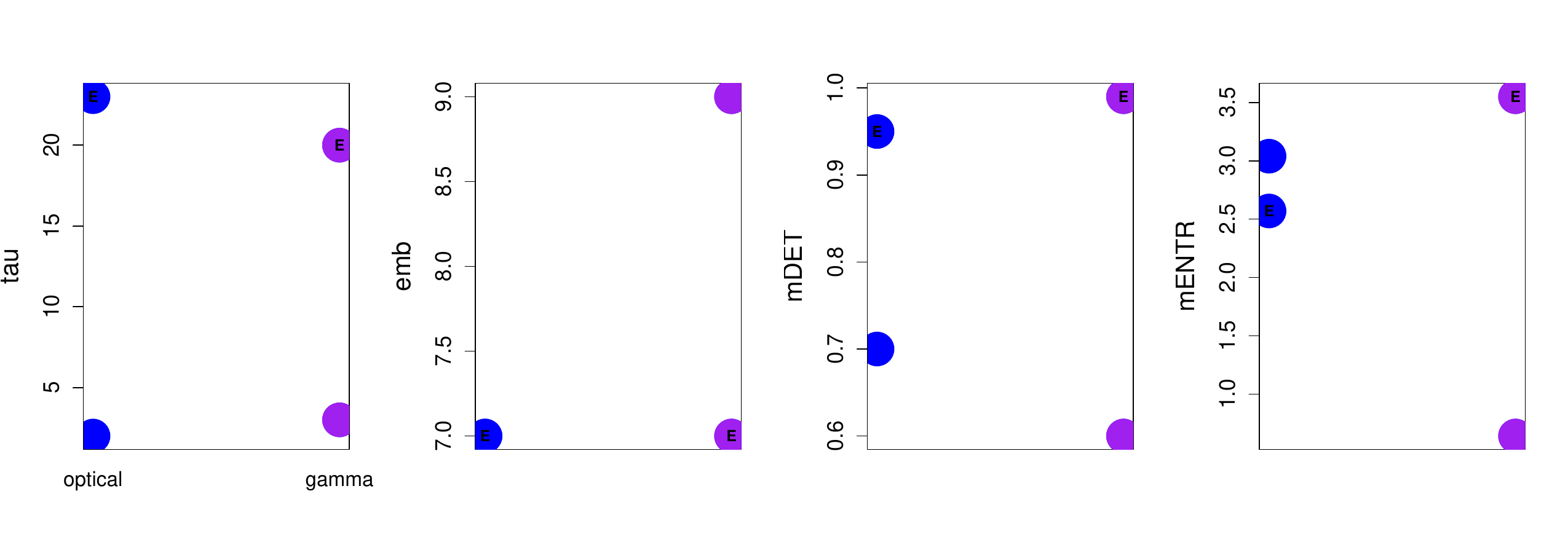}
\caption{This figure displays the quantities from the Tab. \ref{tab}, where blue dots correspond to the optical data measures and purple dots to the $\gamma$-ray data measures.
It can be observed that the difference between the measures of the raw and the artificially evenly binned data (denoted by E) is affected by the amounts of missing/interpolated values. The presence of missing values in the raw data and the inclusion of interpolated values potentially bias the nonlinear measures away from the true values.
}		 \label{ff3}
\end{figure*}

The light curves utilized for nonlinear analysis are depicted in Fig.\,\ref{ff1}, with red indicating points added to achieve uniform sampling from the original light curves. The percentages of added interpolated points are listed in the $inter$ column of Tab.\,\ref{tab}. Both the optical and $\gamma$-ray light curves exhibit significant irregularity and sparse sampling due to the  observational constraints. Consequently, the calculated nonlinear measures, particularly the embedding parameters, are affected, resulting in significant discrepancies between the measures of raw and evenly binned light curves, as shown in Tab. \ref{tab} and Fig.\,\ref{ff3}.

The time lag estimate measure  shows quite big discrepancies between the raw and interpolated data. It is therefore not very clear to determine the lagging behaviour in the emission. However, the values in both, raw and interpolated cases, are close to each other, suggesting similar lagging behaviour for both spectra.

The dimensionality estimate of the system, on the other hand, shows a moderate deviation, slightly suggesting that $\gamma$-ray emission might reflect more complex dynamics. In most cases, both spectra show the same value of 7.

The RQA measures when averaged\footnote{It is in the nature of RQA that the quantities belonging to high RRs contribute more to the final number. This means that these numbers should not be taken literally, for example, when evaluating the deterministic content in a light curve.} and calculated with the application of embedding show also discrepancies for interpolated and raw data apart of the case of the optical emission,  in entropy estimate. Also, the estimate of  determinism in optical data seems more reliable opposite to the $\gamma$-ray data, indicating quite high amount. These results suggest that the optical data likely originate from a complex deterministic system. The measures for $\gamma$-ray data indicate a similarly high deterministic and complex system in the interpolated case. However, in the raw data case, they suggest the opposite—a lower deterministic (more stochastic) ordered system. This difference could be attributed to observational constraints, making it difficult to draw any firm conclusions.

\begin{figure*} 
		\centering 
		\includegraphics[width=1\linewidth]{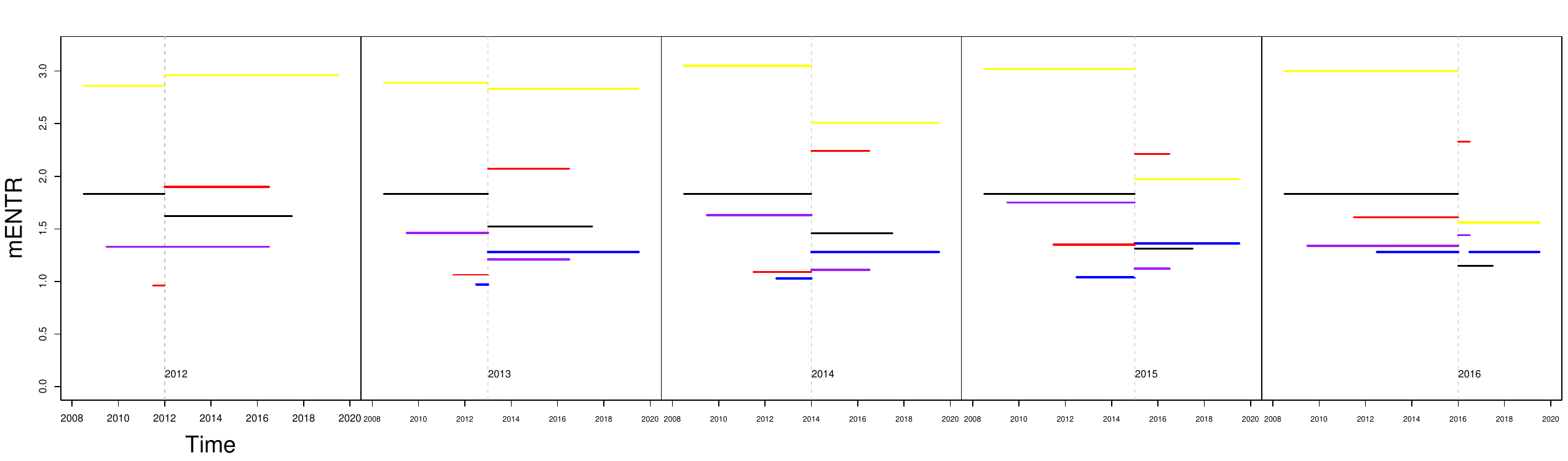}
		
\includegraphics[width=1\linewidth]{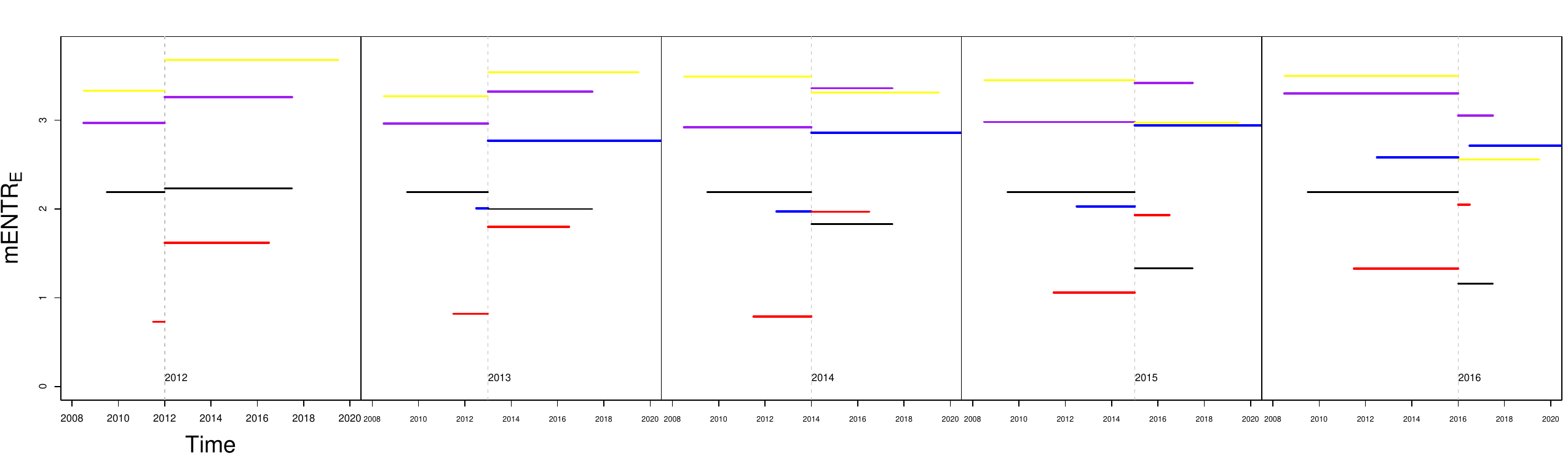}		
\caption{This figure points to potential changes in the jet dynamics. We divided the light curves into two segments over the years and performed an RQA entropy analysis on each segment. Our focus was on identifying changes in system dynamics at transition points indicated by dashed horizontal lines. The entropy values are depicted constant within specific time periods. The placement of horizontal lines indicates the level of system orderliness, with lower placed lines suggesting more ordered systems and higher lines indicating greater complexity in the observed time period.
Gaps between lines of the same color signify shifts in the dynamics.  The blue colour denotes the optical light curve of PKS\,1717$+$177, and purple the $\gamma$-ray data of PKS\,1717$+$177. For better  comparison we added also the measures of  previously analysed source  PKS\,1502$+$106 from \citet{britzen_1502} and also  an artificial light curve. 
Black lines denote the $\gamma$-ray data of PKS\,1502$+$106, and yellow lines denote radio data of PKS\,1502$+$106. The red colour denotes here the artificially generated light curve, which has until 2014 a strong signal (deterministic) content and is almost pure red noise (stochastic) after 2014. It is observed that PKS\,1717$+$177 generally transitions from a more ordered to a less ordered system, similar to the behavior of the artificial light curve, as the horizontal line on the left of the brake, is most of the time lower as the line on the right. This trend is observed for both the raw and interpolated optical data of PKS\,1717$+$177 (blue line), as well as in the interpolated $\gamma$-ray data (purple line).}		 \label{ff2}
\end{figure*}

For the study of the potential sudden change in the jet dynamics, we created plots where we gradually divided the light curves into two parts over the course of several years, see Fig.\,\ref{ff2}. Subsequently, we applied RQA measure of averaged entropy to both segments of the light curves to identify the potential transition time region in the dynamics of the studied system. However, nonlinear measures are typically strongly dependent on the length of the examined time series, especially so in the case of short time series. Thus, the given segmentation relevance is primarily higher when the analysed divided time series have approximately equal lengths. For that fact the segmentation was done only for 5 years from 2012-2016 and the entropy measure was calculated without application of embedding parameters.

For comparison of dynamics we added to this analysis also the measures applied on data from  previously analysed source  PKS\,1502$+$106  \citep{britzen_1502} and also  an artificial light curve to demonstrate the ability of this analysis to capture the shift in the dynamics.  In Fig.\,\ref{ff2}, black lines denote the measures of $\gamma$-ray data of PKS\,1502$+$106, and yellow lines denote the measures of radio data of PKS\,1502$+$106. 
The measurements of the artificial light curve are shown in red. It was generated using the R package RobPer \citep{thieler2016robper}. The first half consists of a simply generated (deterministic) light curve, while the second half is almost pure red noise, indicating a stochastic process. To mimic the missing values in the artificial light curve, $30 \%$ of the data were randomly deleted to simulate missing values. However, there are a number of other factors, currently unknown, that would need to be included in the artificial data to mimic better the observed dynamic properties.  These factors likely affect the entropy estimate values, but the shift in dynamics from deterministic to stochastic is clearly indicated by the gap between the horizontal lines. In Fig.\,\ref{ff2}, the biggest gap between the red line parts is present in 2014, which exactly brakes the artificial light curve into parts with deterministic and stochastic content, while the stochastic end part was all the time denoted with higher entropy.
This kind of behaviour is mostly present in the optical (blue) emission in both raw and interpolated data of PKS\,1717$+$177, suggesting some kind of transition around the year 2015.  The $\gamma$-ray emission (purple) does not exhibit much similarities in raw and interpolated data, but the mean entropy difference and the increase trend are also present in the interpolated $\gamma$-ray data.

Interestingly, the other type of transition is present in the data of the already studied source PKS\,1502$+$106, where the radio (yellow) and $\gamma$-ray (black) emission exhibit a transition from less ordered (complex) to more ordered as there is possible to observe the decrease of entropy when splitting the light curves.

\begin{table*} 
\centering
\begin{tabular}{lrrrrrrrrr}
  \hline
$Obs$ &  $inter$ & $tau$ & $emb$ & $tau_E$ & $emb_E$ & $mDET$ & $mENTR$ & $mDET_E$ & $mENTR_E$ \\ 
  \hline
 PKS\,1717+177 optical &  75 & 2 & 7 & 23 & 7 & 0.70 & 3.04 & 0.95 & 2.57 \\ 
PKS\,1717+177 $\gamma$-rays (\textit{Fermi}) &  89 & 3 & 9 & 20 & 7 & 0.60 & 0.64 & 0.99 & 3.55 \\ 
   \hline
\end{tabular}\caption{This table shows the calculated measures for particular light curves of PKS\,1717+177, the "inter" column denotes the amount of interpolated points to achieve even light sampling, "tau" - time lag, "emb" -embedding dimension, "mDET" - mean determinism and  "mENTR" - mean entropy. The lower "E" index denotes the quantities for interpolated data.   \label{tab}
}	
\end{table*}

\bsp	
\label{lastpage}
\end{document}